\documentclass{emulateapj}

\usepackage{amsmath}

\newcommand{\tnm}{\tablenotemark}
\newcommand{\tnt}{\tablenotetext}

\newcommand{\flux}{ergs cm$^{-2}$ s$^{-1}$}
\newcommand{\intens}{ergs cm$^{-2}$ s$^{-1}$ deg$^{-2}$}

\newcommand{\cdens}{cm$^{-2}$}

\newcommand{\xmmnewton}{{\it XMM-Newton}}
\newcommand{\chandra}{{\it Chandra}}
\newcommand{\asca}{{\it ASCA}}
\newcommand{\rxte}{{\it RXTE}}
\newcommand{\rosat}{{\it ROSAT}}
\newcommand{\bepposax}{{\it BeppoSAX}}

\newcommand{\logn}{$\log{N}/\log{S}$}

\newcommand{\im}{\item}

\newcommand{\dgr}{$^{\circ}$}

\begin{document}

\title{Absolute measurement of the unresolved cosmic X-ray background \\ in the 0.5--8 keV
 band with {\it CHANDRA}}
\shorttitle{UNRESOLVED COSMIC X-RAY BACKGROUND}
\shortauthors{HICKOX \& MARKEVITCH}
\author{Ryan C. Hickox}
\author{Maxim Markevitch\altaffilmark{1}}
\affil{Harvard-Smithsonian Center for Astrophysics, 60 Garden Street,
 Cambridge, MA 02138}
\altaffiltext{1}{Also Space Research Institute, Russian Acad.\
Sci., Profsoyuznaya 84/32, Moscow 117997, Russia}

\slugcomment{Accepted for publication in The Astrophysical Journal}

\setcounter{footnote}{1}

\begin{abstract}
We present the absolute measurement of the unresolved 0.5--8 keV
cosmic X-ray background (CXB) in the \chandra\ Deep Fields (CDFs) North and
South, the longest observations with \chandra\ (2 Ms and 1 Ms,
respectively).  We measure the unresolved CXB intensity by extracting
spectra of the sky, removing all point and extended sources detected
in the CDF. To model and subtract the instrumental background, we use
observations obtained with ACIS in stowed position, not exposed to the
sky.  The unresolved signal in the 0.5--1 keV band is dominated by
diffuse Galactic and local thermal-like emission.  We find unresolved
intensites in the 0.5--1 keV band of $(4.1\pm0.3)\times10^{-12}$
\intens\ for CDF-N and $(5.0\pm0.4)\times10^{-12}$ for CDF-S.  In the
1--8 keV band, the unresolved spectrum is adequately described by a
power law with a photon index $\Gamma=1.5^{+0.5}_{-0.4}$ and
normalization $2.6\pm0.3$ photons cm$^{-2}$ s$^{-1}$ keV$^{-1}$ sr$^{-1}$ at 1
keV.  We find unresolved CXB intensities of $(1.04 \pm 0.14)
\times10^{-12}$ \intens\ for the 1--2 keV band and $(3.4 \pm 1.7)
\times10^{-12}$ \intens\ for the 2--8 keV band.  Our detected
unresolved intensities in these bands significantly exceed the
expected flux from sources below the CDF detection limits, if one
extrapolates the \logn\ curve to zero flux.  Thus these background
intensities imply either a genuine diffuse component, or a steepening
of the \logn\ curve at low fluxes, most significantly for energies
$<$2 keV.  Adding the unresolved intensity to the total contribution
from sources detected in these fields and wider-field surveys, we
obtain a total intensity of the extragalactic CXB of $(4.6 \pm 0.3)
\times10^{-12}$ \intens\ for 1--2 keV and $(1.7 \pm 0.2)
\times10^{-11}$ \intens\ for 2--8 keV.  These totals correspond to a
CXB power law normalization (for $\Gamma=1.4$) of 10.9 photons cm$^{-2}$
s$^{-1}$ keV$^{-1}$ sr$^{-1}$ at 1 keV.  This corresponds to resolved
fracations of $77\pm3$\% and $80\pm8$\% for 1--2 and 2--8 keV,
respectively.
\end{abstract}


\section{Introduction}\label{intro}
Measurement of the cosmic X-ray background (CXB) has been a major
effort in X-ray astronomy since it was first discovered in rocket
flights in the 1960's \citep{giac62}.  The total spectrum of the CXB
has been studied at energies up to 50 keV by {\it HEAO--1} and rocket
experiments \citep[e.g.,][]{mars80, mcca83, garm92, revn04}; for a review of
pre-\rosat\ results, see \citet{mcca90}. It was later studied in
different parts of the 0.5--10 keV interval by \rosat\
\citep[e.g.,][]{snow95, geor96, kunt01}, \asca\
\citep[e.g.,][]{gend95,chen97,miya98,ueda99,kush02}, \bepposax\
\citep[e.g.,][]{parm99,vecc99}, \xmmnewton\ \citep[][hereafter
DM04]{lumb02,delu04} and \rxte\ \citep{revn03}. Deep observations with
\chandra\ were used to study the component of the 0.5--8 keV CXB that
resolves into point sources \citep[e.g.,][and later works]{bran01a,
giac02}. \citet{mark03} used \chandra\ \mbox{ACIS-S} to study the
diffuse components at 0.5--2 keV.  For $E<1$ keV, the CXB is due to
extragalactic and local discrete sources, as well as diffuse (Galactic
and possibly Solar System) components \citep[e.g.,][]{kunt00, crav00}.
Above 1 keV, the CXB is primarily extragalactic in origin, and is well
fit by a power law with a photon index of 1.4.

\defcitealias{more03}{M03}
\defcitealias{delu04}{DM04}

As X-ray telescopes have improved in angular resolution, more and more
of the CXB above 1 keV has been resolved into point sources, mostly
active galactic nuclei \citep[see][for a review]{bran05}.  However
there still remains unresolved CXB flux of unknown origin and
uncertain intensity.
\citet[][hereafter M03]{more03} added the contributions of detected
point sources from a variety of narrow and wide-field X-ray surveys,
and found resolved fractions of the extragalactic CXB of $94\pm7$\%
for the 0.5--2 keV band and $89^{+8}_{-7}$\% for the 2--10 keV band.
\citet{wors04} performed a similar study to find the fraction of the CXB
 that is resolved as a function of energy, using the detected sources
from \chandra\ and \xmmnewton\ observations, and total CXB estimates
from the \xmmnewton\ study of \citetalias{delu04}.  They
found that the resolved fraction of the extragalactic background
decreases significantly at higher energies, from $\sim$80\% at
$\sim$1 keV to $\sim$60\% at 7 keV.  This leaves room for a possible
population of faint sources that have yet to be detected, as well as
truly diffuse components, including exotic ones such as
emission from dark matter particle decay \citep[e.g.,][]{abaz01}.
There remains a great deal of uncertainty in the resolved
fraction, largely due to uncertainty in the absolute flux of the CXB,
and to cross-calibration uncertainties between
different measurements.

Due to its angular resolution, \chandra\ is by far the best instrument
for detecting point sources to very low fluxes and resolving the CXB.
This study uses \chandra\ \mbox{ACIS-I} for an absolute measurement of
the intensity of the {\em unresolved}\/ X-ray background in the energy
range 0.5--8 keV, after the exclusion of sources down to the lowest
fluxes detectable in the deepest current exposures.  We use data from
the \chandra\ Deep Fields \citep[CDFs, e.g.,][]{bran01a,giac02}, which
were designed specifically to resolve as much of the extragalactic CXB
as possible.  As a byproduct of our measurement, we add our unresolved
flux to the contributions from known sources (from the CDFs and other
observations) to obtain the total intensity of the CXB.  This
measurement has not previously been performed with \chandra\ because
of difficulties in determining the instrumental ACIS backgrounds.
Recent calibration using the ACIS detectors stowed out of the focal
plane have made this study possible.  Throughout this paper we will
define the power law photon index as $\Gamma$, where the photon flux
$F\propto E^{-{\Gamma}}$, and we will use 68\% errors.

\section{Data and strategy}\label{obse}
We use two of the deepest observations of the X-ray sky ever performed,
the \chandra\ Deep Fields North and South (CDF-N and CDF-S), which
have total exposure times of $\sim$2 Ms and 1 Ms, respectively.  These
fields are located in regions of low $N_{\rm H}$, away from any
bright features in the Galactic emission.  Each CDF data set
is made up of a number of observations from different epochs covering
approximately the same field.  We use all observations taken after
2000 January 21 when the ACIS focal plane temperature was set at
$-120$ \dgr C; this is the period for which the detector backgound calibration is applicable.  The  observations are listed in Table \ref{tblobs}.

\defcitealias{alex03}{A03}

\begin{figure*}
\plottwo{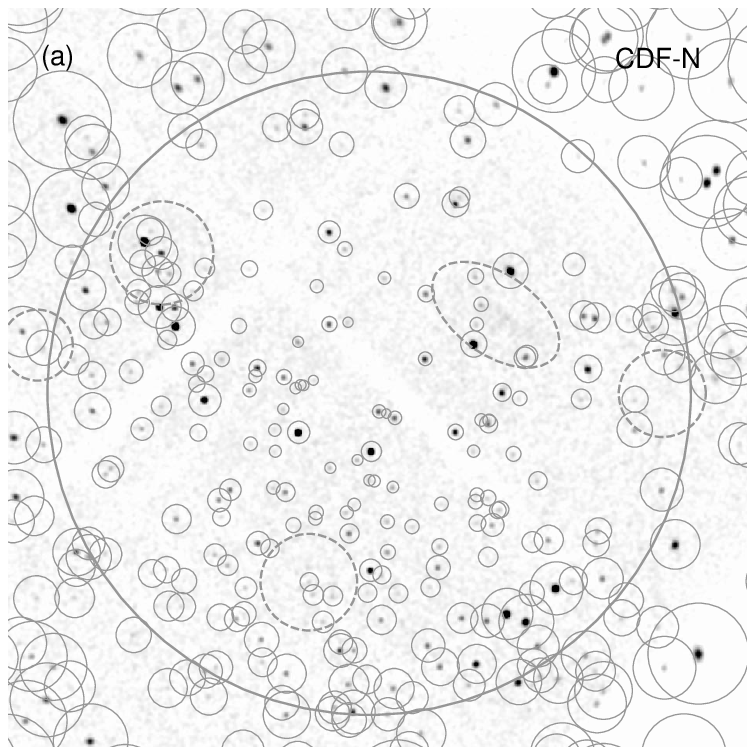}{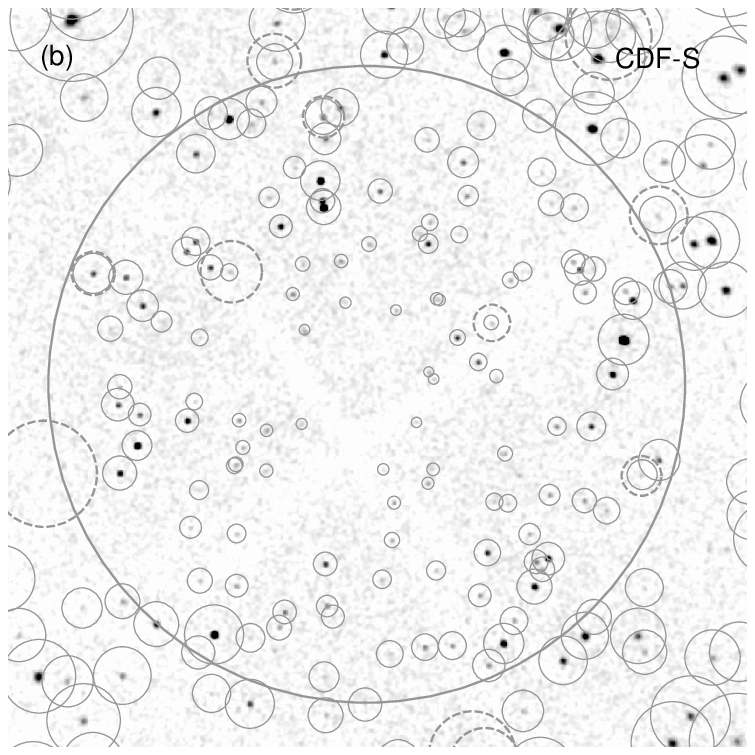}
\caption{Full (1.9 Ms and 0.8 Ms) 0.5--8 keV images for (a) CDF-N and
  (b) CDF-S.  Point source exclusion regions, with source positions
  and fluxes taken from \citetalias{alex03} are shown (for illustration, we
  use regions for the particular aimpoints of ObsIDs 3294 and
  582).  Diffuse source
  exclusion regions are shown as dashed lines. The central 5\arcmin\ radii around the
  aimpoints (the regions used in our analysis) are shown as thick lines.\label{figim}}
\end{figure*}

\begin{deluxetable*}{lrrrrrr}
\tabletypesize{\scriptsize}
\tablecaption{{\it Chandra} Deep Field exposures used for spectral analysis\label{tblobs}}
\tablewidth{16cm}
\tablehead{ 
\colhead{ObsID} &
\colhead{Obs. start date} & 
\colhead{ACIS mode}    &
\colhead{Total exp. (ks)} &
\colhead{Clean exp. (ks)} &
\colhead{RA} &
\colhead{Dec}}
\startdata
\cutinhead{CDF-N}
1671 & 2000-11-21 & F & 168.1 &  97.4 & 12 37 04.30 & +62 13 04.2 \\
2344 & 2000-11-24 & F &  93.3 &   0.0 & 12 37 04.17 & +62 13 03.5 \\
2232 & 2001-02-19 & F & 131.8 & 102.7 & 12 36 35.96 & +62 14 37.4 \\
2233 & 2001-02-22 & F &  64.7 &  37.3 & 12 36 35.66 & +62 14 35.5 \\
2423 & 2001-02-23 & F &  69.8 &  59.1 & 12 36 35.66 & +62 14 35.7 \\
2234 & 2001-03-02 & F & 167.2 & 127.6 & 12 36 34.70 & +62 14 28.8 \\
2421 & 2001-03-04 & F &  63.2 &  47.7 & 12 36 34.39 & +62 14 26.2 \\
Total CDF-N F & \nodata & F & 758.0 & 471.7 & \nodata & \nodata \\
\\
3293 & 2001-11-13 & VF & 161.4 & 113.0 & 12 36 51.77 & +62 13 03.2 \\
3388 & 2001-11-16 & VF &  49.6 &   0.0 & 12 36 51.80 & +62 13 03.5 \\
3408 & 2001-11-17 & VF &  67.7 &  37.3 & 12 36 51.78 & +62 13 03.1 \\
3389 & 2001-11-21 & VF & 125.6 &   0.0 & 12 36 51.70 & +62 13 04.5 \\
3409 & 2002-02-12 & VF &  82.5 &  62.2 & 12 36 36.93 & +62 14 41.1 \\
3294 & 2002-02-14 & VF & 170.8 & 107.8 & 12 36 36.92 & +62 14 41.1 \\
3390 & 2002-02-16 & VF & 164.7 &  94.4 & 12 36 36.92 & +62 14 41.0 \\
3391 & 2002-02-22 & VF & 164.7 & 122.4 & 12 36 36.93 & +62 14 41.1 \\
Total CDF-N VF & \nodata & VF & 986.9 & 537.1 & \nodata & \nodata \\
\cutinhead{CDF-S}
441 & 2000-05-27  & F &  57.1 &  20.7 & 03 32 26.84 & -27 48 17.9 \\
582 & 2000-06-03  & F & 132.2 &  94.3 & 03 32 26.91 & -27 48 16.6 \\
2406 & 2000-12-10 & F &  30.7 &  20.7 & 03 32 28.41 & -27 48 38.3 \\
2405 & 2000-12-11 & F &  60.5 &  30.1 & 03 32 28.88 & -27 48 45.3 \\
2312 & 2000-12-13 & F & 125.3 &  84.0 & 03 32 28.34 & -27 48 38.8 \\
1672 & 2000-12-16 & F &  96.3 &  74.7 & 03 32 28.78 & -27 48 46.3 \\
2409 & 2000-12-19 & F &  70.2 &  55.0 & 03 32 28.09 & -27 48 40.4 \\
2313 & 2000-12-21 & F & 131.8 &  97.5 & 03 32 28.10 & -27 48 40.4 \\
2239 & 2000-12-23 & F & 132.2 &  91.2 & 03 32 28.10 & -27 48 40.3 \\
Total CDF-S & \nodata & F  & 836.5 & 568.0 & \nodata & \nodata \\
\enddata                                                
\end{deluxetable*}

To exclude point sources down to as low fluxes as possible, we will in
effect use the full (2 or 1 Ms) exposures (including several
observations prior to 2000 January 21), by utilizing the source lists
from \citet[][hereafter A03]{alex03}, who performed detailed source
detection for these fields.  For CDF-S, the \citetalias{alex03}
catalog has the same flux limits and is almost identical to the
catalog of \citet{giac02}, although with more accurate source
positions.  For each observation we only use the central 5\arcmin\
radius around the aimpoint, where the source detection flux limits are
lowest and the \chandra\ point-spread function (PSF) has a radius
$\simeq$1\arcsec--3\farcs5, narrow enough to effectively separate source
and background photons.

The main difficulty in such studies with ACIS (and with \xmmnewton,
for that matter) is subtracting the detector background, which
consists of quiescent and flare components.  As we will see in this
paper, the quiescent background is $\sim$5 and $\sim$25 times larger
than the unresolved sky signal for the 1--2 keV and 2--8 keV bands,
respectively, and so requires very careful subtraction.  We model the
quiescent background using the ACIS stowed background calibration,
which was performed from 2002 to 2005.

The CDF data were taken with \mbox{ACIS-I}, which is better suited
than ACIS-S for absolute CXB measurements, due to a lower and much
more stable detector background.  At our present level of
understanding, difficulty in removing the detector background flares
makes this study impossible with \mbox{ACIS-S} \citep{mark03}, but it
is possible to clean flares out to very high precision for
\mbox{ACIS-I}.  To remove low-level flares, we use time filtering with
much stricter criteria than those normally used for extended source
analysis, discarding 39\% of the total CDF exposure (see \S\
\ref{flarback}).  Since the ACIS stowed dataset is currently only 236
ks long, this does not limit our accuracy which is dominated by the
statistical and systematic uncertainties of the stowed dataset.

For our measurements we divide the CDF data into three subsets: the
more recent CDF-N observations taken in Very Faint (VF) ACIS mode (see
\S\ \ref{vfmode}), the earlier Faint (F) mode CDF-N observations, and
the CDF-S data, also in F mode, hereafter CDF-N VF, CDF-N F, and
CDF-S.  The most reliable measurement in the 2--8 keV band comes from
the CDF-N VF subset, which were taken within a year of the earliest
stowed observations, after which we can say with reasonable confidence
that the quiescent background did not change (\S\ \ref{consspec}).  We
will find, however, that the two earlier datasets give results in good
agreement with CDF-N VF, so we will average all three.

\section{Data preparation}\label{specextr}
For each observation, our processing of the X-ray event lists is
almost identical to the standard CIAO pipeline Level 2 processing.
The only minor difference is that in removing bad columns, we
do not also exclude adjacent columns as in the standard pipeline.

\subsection{Coordinate registration}
For the purpose of this work we require the individual exposures of
each of the CDF fields to be aligned as accurately as possible.  We
register the reference frame of each exposure by first detecting point
sources using a wavelet detection algorithm \citep{vikh98}.  We then
translate the images to align the brightest 20--50 detected sources
over the entire field of view, with 0.5--8 keV fluxes between $5\times10^{-4}$
and 0.02 counts s$^{-1}$, to the corresponding RA, Dec positions in
the \citetalias{alex03} catalog.  The \citetalias{alex03} positions
were themselves registered using comparison to accurate radio
positions.  We verify the accuracy of our registration by eye for
each observation.  Simple translation of the images, by not more than
2\farcs6 for any observation, is sufficient to register the positions
to better than $0\farcs 5$.

\subsection{Source exclusion}\label{sourexcl}
For each of the exposures in Table \ref{tblobs}, we extract spectra of
the sky excluding point and diffuse sources.  We use the catalog of
\citetalias{alex03}, after running our own source detection (described
later in \S\ \ref{faintsource}) and obtaining essentially identical source
catalogs.  The \citetalias{alex03} flux limits at the aimpoint for
CDF-N are $2.5\times10^{-17}$ \flux\ and $1.4\times10^{-16}$ \flux\
for 0.5--2 and 2--8 keV, and for CDF-S are a factor of 2 higher.  We
include only the central 5\arcmin\ around the aimpoint, because at
greater off-axis angles the \chandra\ PSF becomes too large for our
purposes.  Around each point source we define a circular exclusion
region.  An estimate for the 90\% encircled energy radius (at $E=1.5$
keV) as a function of the off-axis angle $\theta$ is given
approximately by\footnotemark:
\begin{equation}
r_{90}\simeq 1\arcsec+10\arcsec(\theta/10\arcmin)^2.
\end{equation}
 \footnotetext{\chandra\ Proposer's Observatory Guide (POG) v7.0,
 Fig.\ 4.13, available at
 {\tt http://cxc.harvard.edu/proposer/POG}.} 
To be sure to fully exclude source photons from the wings of the PSF,
we multiply $r_{90}$ by a factor that varies depending on the flux of
the excluded source.  We find that it is sufficient to use exclusion
radii of 4.5, 6, and 9 times $r_{90}$ for sources with $<100$,
$100-1000$, and $>1000$ total source photons, respectively, in the
0.5--8 keV energy band.  Because
the aimpoints of the observations differ by up to 4\arcmin, we
define the exclusion regions separately for each observation.  Using a
model for the $\chandra$ PSF (from the \chandra\ CALDB), we find that after
such source exclusion, the ``missed'' source flux due to PSF scattering is
$<$0.1\% of the total source flux.  This will correspond to only
$\lessapprox$5\% of the unresolved background intensity (\S\ \ref{psfscatter}).

We also exclude detectable extended sources.  For CDF-N we use the
diffuse source regions given in Table 1 of \citet{baue02}, multiplying
the dimensions by a factor of 1.5 to ensure that diffuse source
photons on the edges are excluded.  For CDF-S we use the sources in
Table 5 of \citet{giac02}, and multiply the FWHM values by 5 to obtain
the exclusion radii.  The surface brightness of these extended sources
is 2--3 times lower than the unresolved CXB brightness that we will
obtain, so the details of source exclusion regions should not be
important; we verify this assumption in \S\ \ref{faintsource}.  In
Fig.\ \ref{figim} we show images with exposures of 1.9 Ms and 0.8 Ms
(after cleaning) of the CDF-N and CDF-S fields for 0.5--8 keV, with
the point source and extended source exclusion regions.

\subsection{VF mode background filtering}\label{vfmode}
Seven of the CDF-N exposures (Table \ref{tblobs}) were taken in VF ACIS
telemetry mode, for which the detector background can be reduced
significantly by rejecting events with signal above the split
threshold in any of the outer pixels of the 5 x 5 pixel event island,
after an approximate correction for the charge transfer inefficiency \citep[CTI,][]{vikh01}.  This
makes the instrumental background lower (by $\sim$20\% for the
2.3--7.3 keV band for the front-illuminated (FI) chips such as
\mbox{ACIS-I}) more spatially uniform, and more stable (see \S\
\ref{consspec}). This filtering makes the CDF-N VF observations the most
reliable for measurements for $E>2$ keV.  The stowed dataset was taken
in VF mode, so the same additional filtering is applied when
appropriate.

The CDF-N VF data were taken with an onboard upper telemetry cutoff of
3025 PHA channels.  This means that any events with a PHA\_RO (the
value of PHA before gain correction) $>$ 3025 ADU, corresponding to
$\sim$12 keV, will not appear in the event lists. Therefore some
events with energies $<$ 12 keV after gain correction, but with
PHA\_RO $>$ 3025, will be missing from the observed 9--12 keV count
rate.  Because we use the 9--12 keV count rate to normalize the stowed
background, which does not have such a cutoff (\S\ \ref{backnorm}),
this effect may impact our results.

We tested for the effects of the PHA cutoff by examining telemetered
data that does not have such a cutoff, including some sky observations
as well as the stowed exposures.  In all cases, a cut of PHA\_RO $<$
3025 produces a 1\% difference in the 9--12 keV count rate for the
full \mbox{ACIS-I} field of view. However, the missing events are
found near the edges of the I-array, because the CTI correction, which
moves energy upwards, is less strong there.  The difference is
negligible in the central 5\arcmin\ region for which we extract
spectra and normalize the background. Therefore, the upper telemetry
cutoff has a no significant effect on our results.

\section{Instrumental background}\label{instback}
 
The total observed background spectrum for \mbox{ACIS-I} consists of
four separate components: (1) the real cosmic background signal,
(2) quiescent and (3) flaring detector backgrounds due mainly to particles of different energies, and (4) a readout artifact from the sources in
the field of view due the finite
readout time of the ACIS detectors, which for our purposes can be
treated as a background.
Accurate removal of the detector backgrounds is the key aspect of this
analysis, so we discuss this in detail.

\subsection{Quiescent background}\label{quieback}
The quiescent instrumental background of the ACIS detectors is due to
interactions of the CCDs and surrounding materials with high-energy particles, and has been measured first by
studying data taken in Event Histogram mode (that does not
telemeter imaging information) with ACIS stowed inside the detector
housing \citep{bill02}, later by briefly observing the dark Moon, and finally
by taking long exposures in full imaging mode with ACIS stowed.  For the
latter measurement, the ACIS position inside the
housing was selected to minimize the flux from the internal
calibration source spectral lines. The satellite was in a regular
region of the orbit where science observations are performed (i.e.,
not in the radiation belts) during these observations.  The detector housing blocks celestial X-rays and
low-energy particles that cause flares (\S\ \ref{flarback}) but does not
affect the quiescent background rate by any detectable amount, as we shall see from comparison with the dark Moon data (\S\ \ref{blockcomp}).  

Thus the ACIS-stowed dataset, which we use here, accurately represents
the quiescent background in ACIS sky exposures.  As of fall 2005, a
total of 236 ks of stowed background data are available, accumulated
during 2002--2005\footnotemark.  Due to telemetry limitations, the stowed data were taken only with
chips I0, I2, I3, S1, S2, and S3; the stowed background for the I1
chip is a slightly scaled and reprocessed copy of the I0 data.  As we
discuss in \S\ \ref{vari1}, this has no significant effect on the
results.

\begin{figure}
\epsscale{1.7}
\plotone{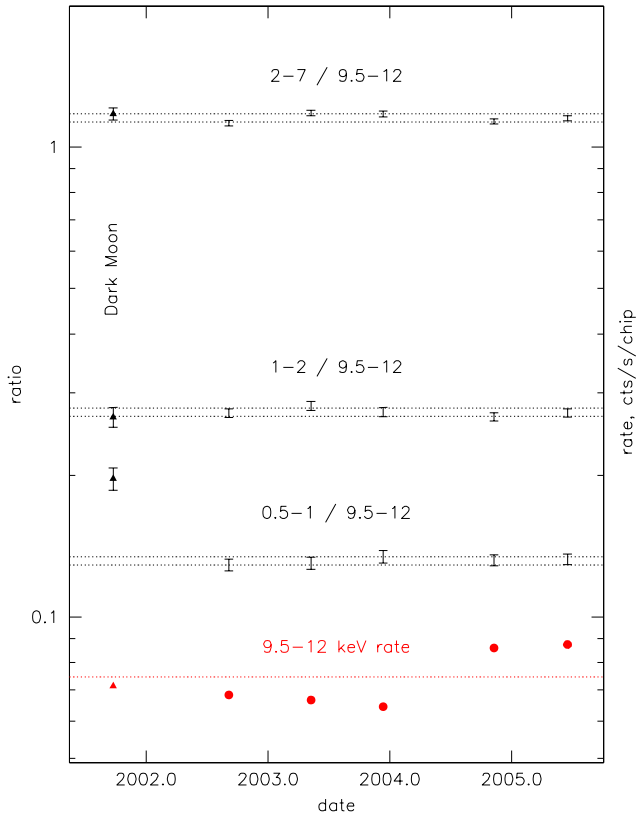}
\caption{Ratios of count rates in various bands to the 9.5--12 keV
  count rate (shown at bottom) for \mbox{ACIS-I} dark Moon (triangles)
  and stowed observations.  Pairs of lines are $\pm$2\%.  A deviation
  in the 0.5--1 keV Moon data is an astrophysical signal
  \citep{warg04}.  \label{figstow}}
\end{figure}

\begin{figure}
\epsscale{1.2}
\plotone{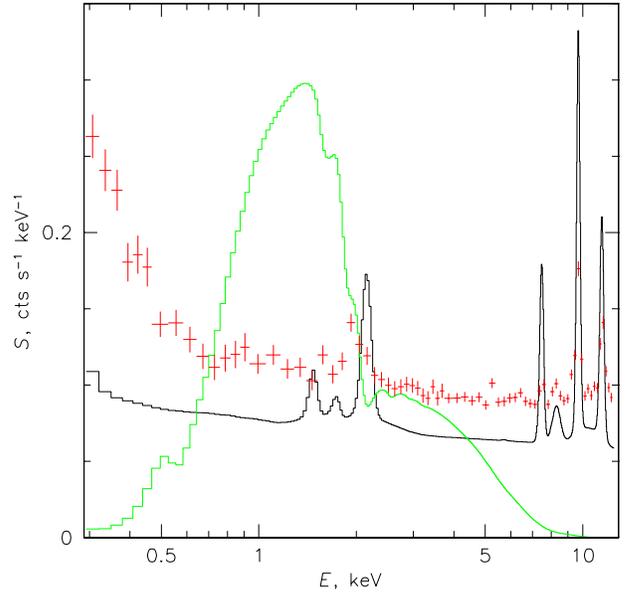}
\caption{Spectral shape of the \mbox{ACIS-I} quiescent background
  (solid dark line), along with a typical power-law sky
  spectrum (solid gray) and the average spectrum of the most common
  \mbox{ACIS-I} flare 
  species (crosses). The flares and quiescent background are normalized to
  have the same 9--12 keV count rate. \label{figflare}}
\end{figure}

\footnotetext{Details on the stowed exposures are given at {\tt http://cxc.harvard.edu/contrib/maxim/stowed/}.}

The overall intensity of the quiescent background varies at the 10--20\%
level (showing a correlation with the solar cycle), but its spectral
shape is remarkably stable.  Fig.\ \ref{figstow} shows the ratios of
rates in three interesting bands to that in the 9.5--12 keV band
(dominated by the fluorescent gold lines, see Fig.\ \ref{figflare}), for
separate $\sim$50 ks exposures included in the ACIS-stowed dataset,
and for the 14 ks dark Moon dataset.  

\subsubsection{Absence of any blocked component in the stowed background}\label{blockcomp}

The first important thing to note from Fig.\ \ref{figstow} is that the
dark Moon spectrum is the same as the stowed background for $E>1$
keV; the 2--7/9.5--12 keV rate ratios are consistent for the two
datasets. For chips I2 and I3, the ratio for the dark Moon is
$1.01\pm0.03$ times that for the corresponding stowed data (note that
in Fig.\ \ref{figstow}, chips I2 and I3 for the Moon are compared to chips I0, I2,
and I3 for the stowed data, hence a small difference). From a separate
Moon observation taken 2 months earlier, the ratio for chip S2, also
an FI chip, is $0.99\pm 0.04$ times that for the S2 stowed data (not
used in this analysis). The average of these factors is $1.00\pm
0.02$, which indicates that any possible quiescent sky background
component not present in the stowed background is essentially zero for
FI chips.  For S3, a back-illuminated (BI) chip which is more strongly affected by
flares, the Moon 2--7/9.5--12 keV ratio is $1.04\pm0.03$ times the
stowed ratio,
consistent with low-level residual flares present in the S3 sky data.
The same is true for the 1--2 keV band, where the ratio for the Moon
data is $0.97\pm0.05$ times that for the stowed background for chips
I2 and I3, $0.97\pm0.06$ for chip S2, and $1.01\pm0.04$ for chip S3.

\begin{figure*}
\epsscale{1.1}
\plotone{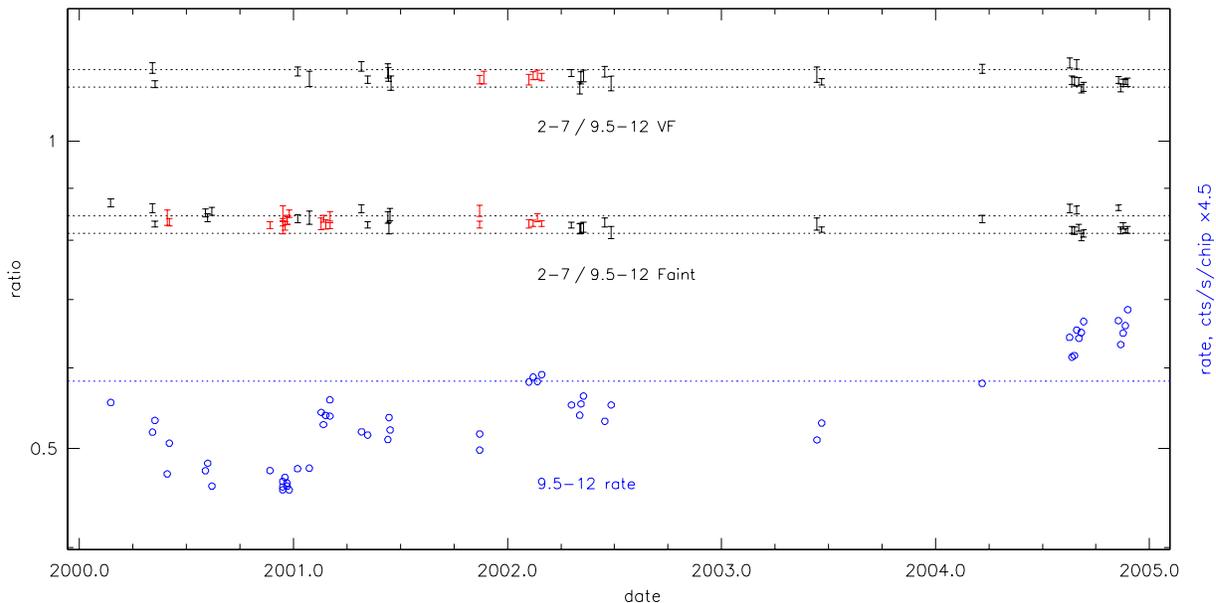}
\caption{Similar plot to Fig.\ \ref{figstow} but for sky observations
  included in the blank-sky datasets, approximately corrected for the
  unresolved sky signal.  The top points show the ratio with VF
  cleaning, the middle points with no VF cleaning, circles show the
  arbitrarily scaled 9.5--12 keV count rate.  The pairs of dotted
  lines show $\pm$2\% around the averages for 2001--2004.  Dates of
  some points are shifted for clarity.  Red points show the CDF
  observations.  For CDF-N VF data, the ratios were corrected for the
  1\% effect of the upper 3025 ADU PHA cutoff (\S\ \ref{vfmode}). Note
  that we use the energy range 9--12 keV for normalizing the
  background spectrum (\S\ \ref{backnorm}), here we show the 9.5--12 keV
  band, for applicability to analysis of fields including brighter
  sources, which can have significant sky flux in the range 9--9.5 keV.  \label{figblank}}
\end{figure*}

\subsubsection{Constancy of the spectral shape}\label{consspec}
Another important fact seen from Fig.\ \ref{figstow} is that while the
total background rate is very different (lower data points), the
ratios stay within a $\pm$2\% rms, some of which is statistical
scatter.  The exception is an excess at $E<1$ keV from the dark Moon,
which is an astrophysical signal \citep{warg04}.  The same conclusion
on the constancy of the background spectral shape within $\pm$2\% rms
was reached from looking at variability of the Event Histogram mode
data \citep{mark03}.

The comparison of a larger number of sky observations included in the
ACIS blank-sky background datasets\footnotemark, which span the
2000-2004 period, also shows an rms scatter of not more than 2\% in
the 2--7 keV to 9.5--12 keV background flux ratio (Fig.\
\ref{figblank}). The observations included in this plot are cleaned of
background flares (using a less rigorous criterion than we will use in
the work below, see \S\ \ref{flarback}), and the detectable point sources
are removed. Because of different exposures (30--160 ks), they include
different contributions of unresolved CXB, which we approximately
subtract using the results obtained in \S\ \ref{faintsource} (Fig.\ \ref{figflux}) and
assuming a uniform CXB over the sky, before dividing by the 9.5--12
keV rate.  This correction is small in the 2--7 keV band but at lower
energies the residual sky flux is too great, so we can only use these observations
for checking the detector background variability at high
energies. Most of the observations are in VF mode, so we show the
ratios both for VF-cleaned and uncleaned (i.e., equivalent to F mode)
data. Earlier observations were obtained in F mode so we cannot
apply VF cleaning to them. 

We note that VF-uncleaned ratios may show
some downward trend during 2000 at a 2\% level, which is probably
related to the slowly changing CTI in
the FI chips.  The CTI and time-dependent gain corrections are
calibrated for real X-ray photons registered in the imaging area of
the CCD, and are not corrected for the undamaged frame store area of the
CCD.  For a fraction of the background that originates in frame store,
these corrections result in energy shifts, thus we may expect apparent
changes of spectral shape with time. The CDF-S and CDF-N F data were
obtained during the above period of a possible trend.  Our systematic
uncertainty will include this, and we will see that the results are
consistent within our systematic uncertainties between all datasets.

\footnotetext{{\tt http://cxc.harvard.edu/contrib/maxim/acisbg/}}

The VF cleaning removes a relatively larger fraction of the
background events originating in frame store, thus reducing these time
dependencies.  Indeed, as seen in Fig.\ \ref{figblank}, the spectral shape of the VF-cleaned
background does not show any trends with time; all ratios are within
the 2\% rms scatter.  This is in contrast to the highly variable
\xmmnewton\ background spectra \citep[e.g.,][]{delu04, neva05}.  Thus we are safe to use the 2002--2005 ACIS-stowed
background to model CDF-N VF observations from the end of 2001 to early
2002.  We will therefore treat the CDF-N VF subset as the most
reliable for background-sensitive 2--8 keV measurements, although we
will see that the other datasets give consistent results.

\subsubsection{Background normalization}\label{backnorm}

Since the \chandra\ effective area in the 9--12 keV band is
negligible, essentially all the 9--12 keV flux in the sky data is due
to particle background (in the absence of event pileup, which is true
for all the CDF observations, see \S\ \ref{readback}).  Because of the
stability of the quiescent background spectral shape, we can therefore
scale the normalization of the stowed spectrum to the observations by
equating the 9--12 keV count rates, to get an accurate model of the
non-sky quiescent background.  In addition to the statistical
uncertainty, we include a $\pm2$\% variation on this normalization to represent
a systematic uncertainty of the background spectral shape inherent
in such modeling, and propagate it into our final results.  This
variation is considered independent between the three CDF datasets
because they are well separated in time (see Table \ref{tblobs}), but is
conservatively taken to be the same for observations within each
dataset, which were usually taken back to back.  This
simple step is adequate and will be much more important for our 2--8 keV
result than for the $E<2$ keV results.  We note that this 2\% background
error may be somewhat conservative for the CDF-N VF data, since the
shape of the VF background spectrum is very stable.

\begin{figure*}
\epsscale{1.1}
\plotone{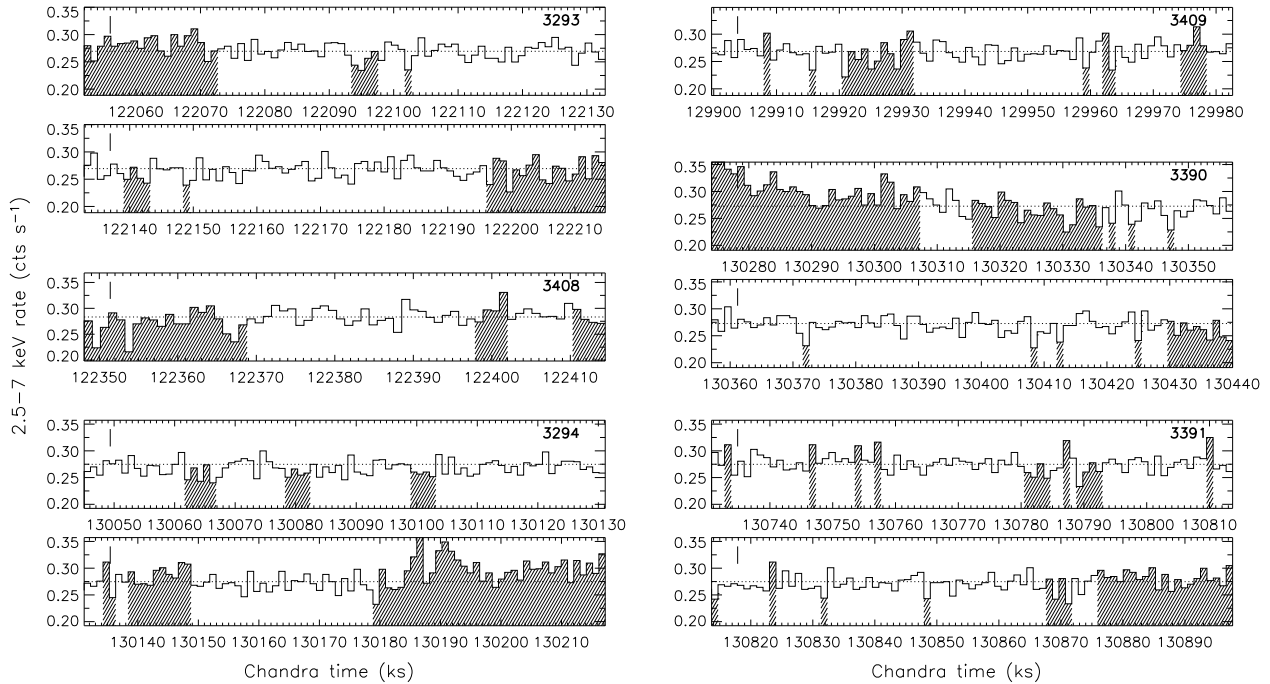}
\vskip-2.0cm
\caption{\footnotesize 2.3--7.3 keV light curves for the
  CDF-N VF observations.
  Bins are 1 ks, count rate errors are shown at upper left.  Dotted
  lines show the mean count rates after cleaning. Shaded bins are excluded by the
  combination of
  cleaning techniques described in \S\
  \ref{flarback}.  Shaded intervals that are smaller than the 1 ks
  bins are due to gaps in good time intervals from standard processing.\label{figlcnvf}}
\end{figure*}

\setcounter{figure}{4}

\begin{figure*}
\epsscale{1.1}
\plotone{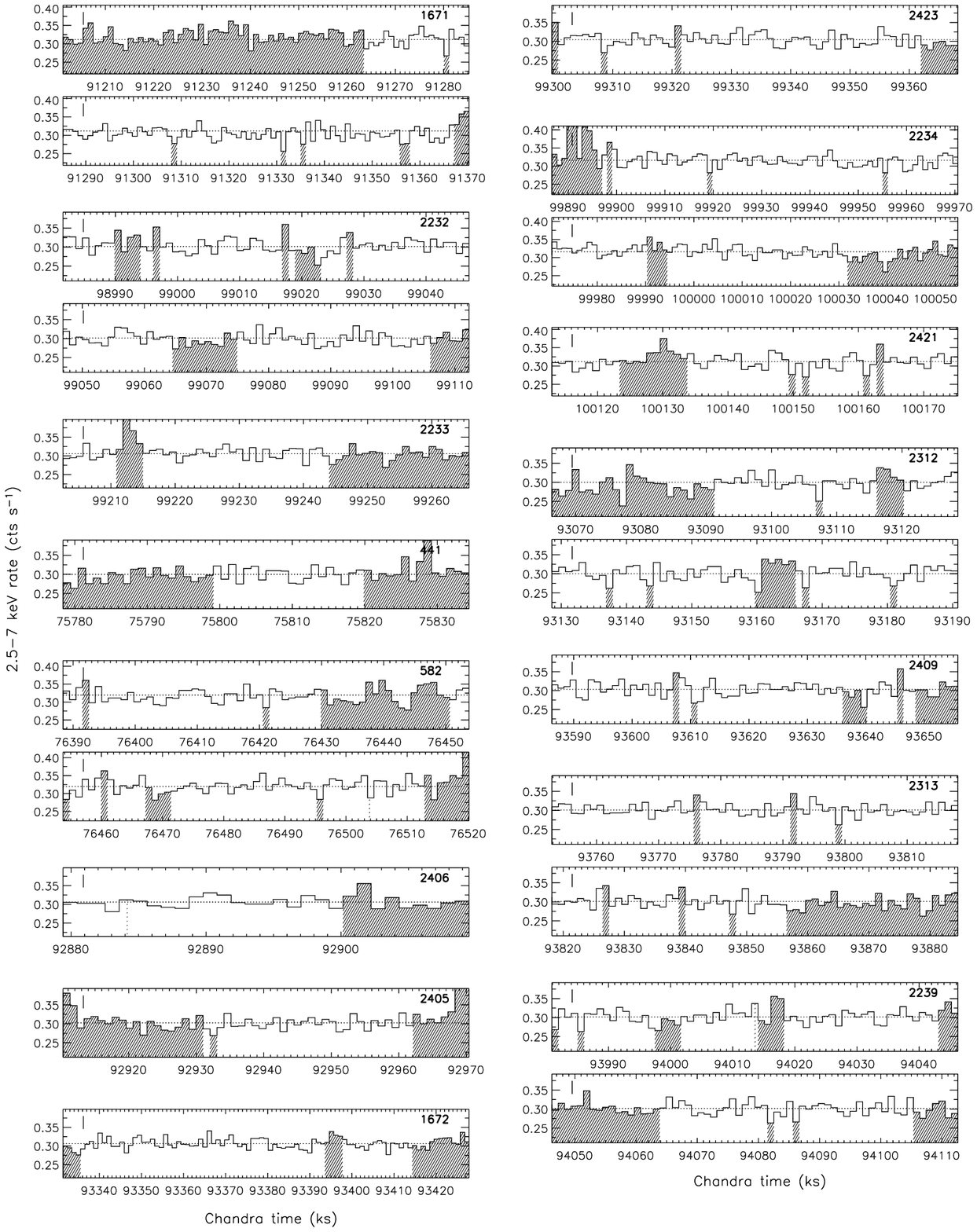}
\vskip-3.1cm
\caption{(contd), for the CDF-N F and CDF-S observations.}
\end{figure*}

\subsection{Flaring background}\label{flarback}
ACIS is also subject to background flares whose spectra are variable
and different from that of the quiescent background.  Because the
present work relies on the accuracy of the background model at a
2--3\% level, excluding these flares is the most critical aspect.  We identify
flares by their time and spectral variability.  We create light
curves for each observation in the 2.3--7.3 keV band, which is the most
sensitive to flares due to the spectral shapes of the flaring and
quiescent backgrounds (the quiescent background plus the sky signal
has a minimum in this band, while the flares do not, see Fig.\ \ref{figflare}).  For light curve extraction we include the
whole \mbox{ACIS-I} array, but exclude the \citetalias{alex03} point sources as
described in \S\ \ref{specextr}, although with exclusion radii 1/3 the
size, so as to allow for greater background count rates and improved
statistics.  Three CDF-N observations (ObsIDs 2344, 3388, and 3389) were
completely excluded from the spectral analysis because of extensive
flaring activity.

For the remaining observations, we filter the light curves for flares
successively in bins of 1 ks, 4 ks, and 10 ks, to try to balance the
sensitivity to flares and statistical scatter.  For each light curve
we calculate a first approximation mean count rate, by rejecting bins
with rate $>$3$\sigma$ over the mean from the full exposure.  We then
further exclude bins where the count rate deviates from this mean by
20\%, 10\%, and 6\% for the 1 ks, 4 ks, and 10 ks binned light curves,
respectively. This corresponds to 2$\sigma$ deviations for each
binning, so we necessarily exclude some statistical devitions.  We
exclude both positive and negative deviations, even though the latter
are mostly statistical fluctations, so as not to bias negatively the
result by removing only positive statistical fluctuations.

\begin{figure*}
\epsscale{1}
\plotone{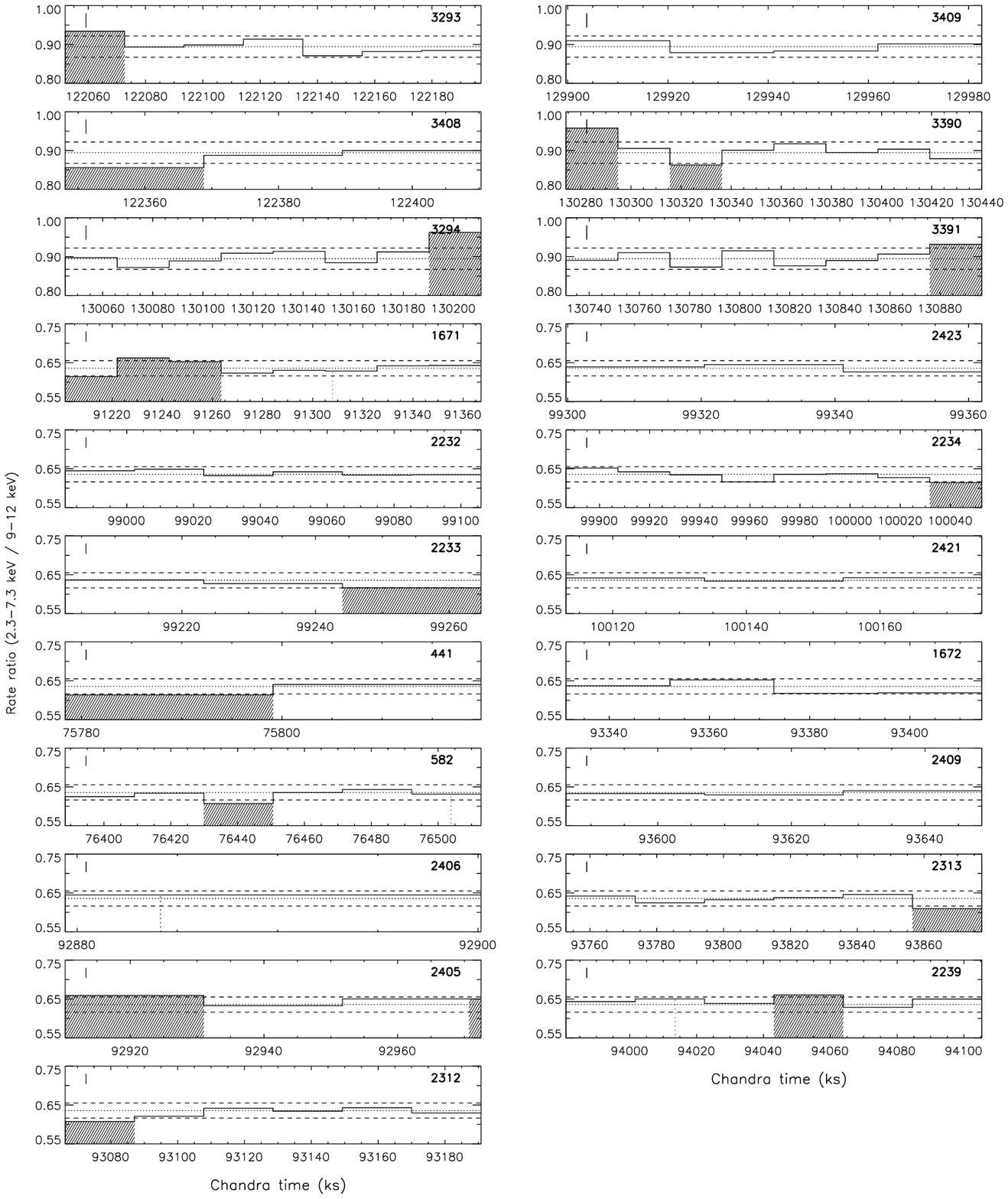}
\vskip-3cm
\caption{Ratio of 2.3--7.3 keV count rate to 9--12 keV count
  rate, in bins of 20 ks, for each observation. $\pm$1$\sigma$
  errors are shown at upper left.  Dotted lines show mean ratios
  after cleaning (note
  the difference between F and VF observations), dashed
  lines show  $\pm$3\ variations.  Time bins excluded by this ratio
  cleaning are shaded, along with periods excluded by the standard
  processing (before our time filtering). \label{figrat}}
\end{figure*}

We also identify flares by using the fact that ACIS flares have
different spectral shapes than the quiescent background.  In
particular, \mbox{ACIS-I} flares often have a power law spectrum with
photon index $\sim$0 for $E>0.5$ keV, and weak or completely absent
fluorescent lines.  Thus for flares, the ratio of 2.3--7.3 keV flux to
9--12 flux will be systematically larger than for the quiescent
background (Fig.\ \ref{figflare}).  We therefore derive the ratio
between the 2.3--7.3 keV and 9--12 keV count rates as a function of time
in bins of 20 ks, and exclude periods with deviations of $>$3\%, which
corresponds to 1.5$\sigma$ statistical scatter.  While the count-rate filtering is performed
using individual mean values for each observation, for this spectral ratio filtering,
we use a single mean value for the nominal ratio (separately for F and VF
data, due to the different background levels).  This ensures that across
all observations the spectral shape is constant.  This kind of
filtering is well-suited to removing low-level flares and has not been
performed before.

Because this ratio filtering includes  the entire \mbox{ACIS-I}
array and not just the central 5\arcmin,
the problem of the upper telemetry cutoff in the CDF-N VF data (\S\
\ref{quieback}) becomes important, in that it may cause a higher
2.3--7.3 keV to 9--12 flux ratio in those observations with more missing
photons having PHA\_RO $>$ 3025.  For the CDF-N VF data, we therefore performed the ratio
filtering again, using the count rate for events with $E>9$ keV and
 PHA\_RO $<$ 3025, which should not be affected by the telemetry cutoff.
This check gives essentially identical output time intervals, and has negligible effect
on the final spectrum.

Background light curves, showing the filtered time intervals, are
shown in Figs.\ \ref{figlcnvf}--\ref{figrat}.  The above filtering
steps, applied separately, exclude 12\%, 14\%, 18\%, and 24\% of the
exposure time for the 1 ks, 4 ks, and 10 ks light curve and 20 ks flux
ratio cleaning, respectively (note that with the larger binning, we
necessarily exclude some time intervals at the start or end of each
observation that are not included in bins). After combining these
filters a total of 32\% of the exposure time
is excluded (39\% if we include the three observations that are
completely excluded due to flares).  This is a much more rigorous cleaning than
normally applied for the \mbox{ACIS-I} extended-source data \citep[which is
limited to our 1ks, 20\% light curve cleaning step, and on average
removes 6\% of the total exposure, e.g.,][]{vikh05}, and is sufficient to remove
flaring intervals to our required accuracy.

\subsubsection{ACIS-I readout artifacts}\label{readback}
A small but significant contribution to the background comes from the
41 ms of time needed to read out the CCD, for each sky exposure of 3.1
or 3.2 s (3.1 s for the majority of the CDF observations).  This leads
to artifacts in the forms of streaks along the CHIPY direction
accompanying all celestial sources\footnotemark. \footnotetext{
     \chandra\ POG, \S\ 6.11.4}  Because these events
lie outside the source exclusion regions, we simulate and remove this
readout background using the {\tt make\_readout\_bg} routine\footnote{{\tt http://cxc.harvard.edu/contrib/maxim/make\_readout\_bg/}} \citep{mark00}. This script takes the original event list and
randomizes the CHIPY values for each event, then recalculates the
photon energies to produce a new event list which approximately
simulates the spectral and spatial characteristics of the readout
artifact.  The method works only when there is no pileup in the
observations.  This is true for both CDF fields; the very brightest
sources in the \citetalias{alex03} catalog have count rates 0.022
(CDF-N) and 0.012 (CDF-S) counts s$^{-1}$, for which pileup will be
$<$3\%\footnotemark.

\footnotetext{\chandra\ POG, Fig.\ 6.18}

Using this readout event list, we extract a spectrum (hereafter the "readout
spectrum") using the same GTI and region filters as for the sky
events.  During spectral fitting, we subtract the readout spectrum
with a normalization of $0.0132=41\:\rm{ms}/3.1\: s$.  This also
subtracts a fraction of the detector background, which is taken into
account when normalizing the stowed dataset.

To summarize, the unresolved sky spectrum is given by
\begin{equation}
f_{\rm unr}=f_{\rm obs}-C_{\rm bg}f_{\rm stowed}-0.0132f_{\rm readout}.
\end{equation}
where $f_{\rm obs}$ is the observed spectrum with sources excluded, and $C_{\rm bg}$ is the stowed background scaling factor, given by 
\begin{equation}
C_{\rm bg}=\frac{f_{\rm{9-12 \: keV} \:  (obs)} -
  0.0132f_{\rm{9-12 \: keV} (readout)}}{f_{\rm{9-12
   \: keV} (stowed)}}.
\end{equation}
By design, this subtraction gives zero flux at 9--12 keV.

\section{Spectral extraction}\label{specanal}

For each observation, we extract source-excluded spectra of the sky,
using the region and time filters described above, specific to each
observation.  Due to differences in pointings of up to 4\arcmin, the exposures in a
given field (and our $r<5$\arcmin\ extraction regions) have slightly
different sky coverages.  Our analysis implicitly assumes that across
the different pointings in a given field, the faint, unresolved sky signal has the
same surface brightness.

For each CDF exposure, we project the stowed events into sky
coordinates using the appropriate aspect solution, using the
{\tt make\_acisbg} routine\footnote{{\tt http://cxc.harvard.edu/contrib/maxim/acisbg/}} \citep{mark00}.  We then extract a spectrum
(hereafter called the "stowed spectrum") using the same region filtering as
for the sky exposures.  The normalization of the background spectrum is
described in \S\ \ref{backnorm}.

For any individual observation, the unresolved signal is too faint for
detailed spectral analysis.  Therefore we created composite spectra
for each of the three subsets of the data, CDF-N VF, CDF-N F, and
CDF-S. The RMFs and ARFs are derived using A. Vikhlinin's ACIS tools
CALCRMF and CALCARF\footnote{{\tt http://cxc.harvard.edu/cont-soft/software/calcarf.1.0.html}}.  These are equivalent to the
standard CIAO tools, except that CALCARF also includes an
area-dependent dead-time correction (a 3\% effect). In each
observation, RMFs and ARFs are averaged over the extraction region.
Between observations, RMFs and ARFs are averaged by exposure time,
using the FTOOLS {\it addrmf} and {\it addarf}\footnote{{\tt
http://heasarc.gsfc.nasa.gov/docs/software/ftools/}}.  The stowed
spectra are also averaged, weighted by the corresponding sky exposure
time.  Note that even though the same stowed background dataset is
used for all observations, the sky coverage is slightly different,
which results in small differences in the stowed background spectra.
Care is taken to treat the background data in a statistically correct
way, and not as independent spectra.  We also approximately take into
account the fact that the I0 and I1 stowed data are not independent
(\S\ \ref{quieback}).  Of the background events 12\% (CDF-N) and 22\%
(CDF-S) are taken from the I1 chip.  We thus increase the statistical
errors on the background spectrum accordingly, by 7\% for CDF-N and
13\% for CDF-S.  The 2\% systematic normalization uncertainty (\S\
\ref{quieback}) is conservatively applied to the combined stowed
spectrum for each dataset.

\subsection{ACIS calibration}\label{aciscal}

The latest calibration is used as described in
\citet{vikh05}. The data processing used the CTI correction CALDB file
ctiN0002, ACIS gain file gain\_ctiN0003, and time-dependent gain
correction t\_gainN0003, the same versions as used for stowed data (the
recently released updates do not affect our results
significantly). The instrument responses included time- and
position-dependent low-energy contamination correction (equivalent to
the CALDB file contamN0004), mirror edge correction function applied
to the CALDB effective area file axeffaN0006, which is equivalent to
using the recently released updated area axeffaN0007, and the dead
area correction to the FI quantum efficiency (QE) by a
position-dependent factor around 0.97.  

Because of these calibration updates, our absolute fluxes will be
slightly different from earlier \chandra\ papers.  In particular,
because of the combined effect of mirror area, dead time and
position-dependent contaminant corrections, our 1--2 keV fluxes should
be within 2--3\% of those in A03, but lower by about 5--7\% at 2--8
keV.  The \mbox{ACIS-S} measurement by \citet{mark03} was
additionally affected by old values of the uncontaminated low-energy
QE for BI chips (not used here), resulting in an overall 7\%
overestimate of the 1--2 keV fluxes. It is difficult to evaluate the
differences with still earlier analyses, e.g., of \citetalias{more03}.

\section{Results}\label{results}

\begin{figure}
\epsscale{1.2}
\plotone{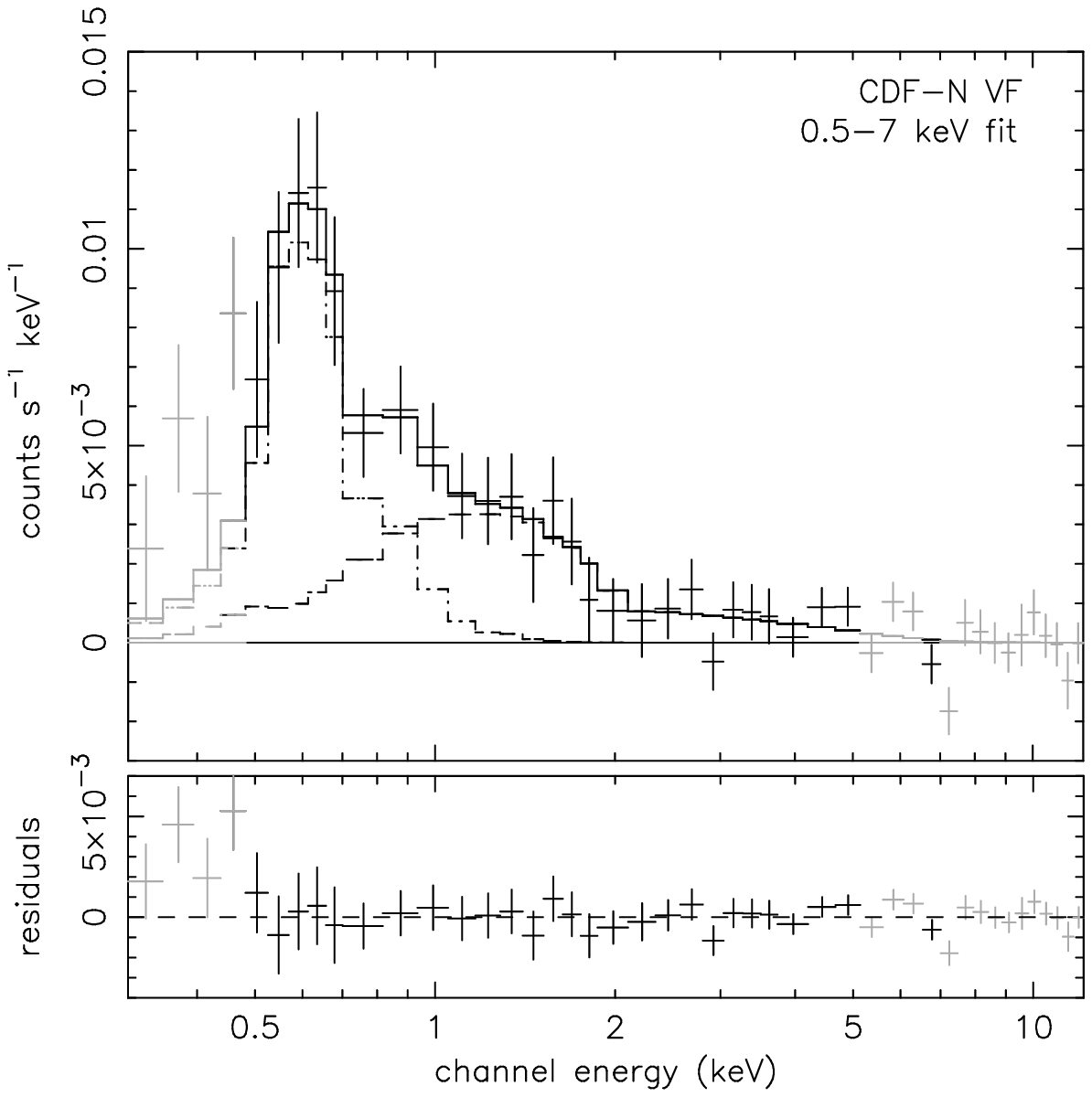}
\caption{Unresolved, background-subtracted sky spectrum for the composite CDF-N VF data.   The
  fit was performed in the energy range 0.5-7 keV (excluding
  5.6--6.2 keV) and consists of a
  power law and APEC components, shown as dashed lines.  We show the
  whole range from 0.3--12 keV for completeness; bins not
  included in the fit are in gray.  Fit results are given in
  Table \ref{tblfits}.\label{figspec}}
\end{figure}

\begin{figure}
\epsscale{1.2}
\plotone{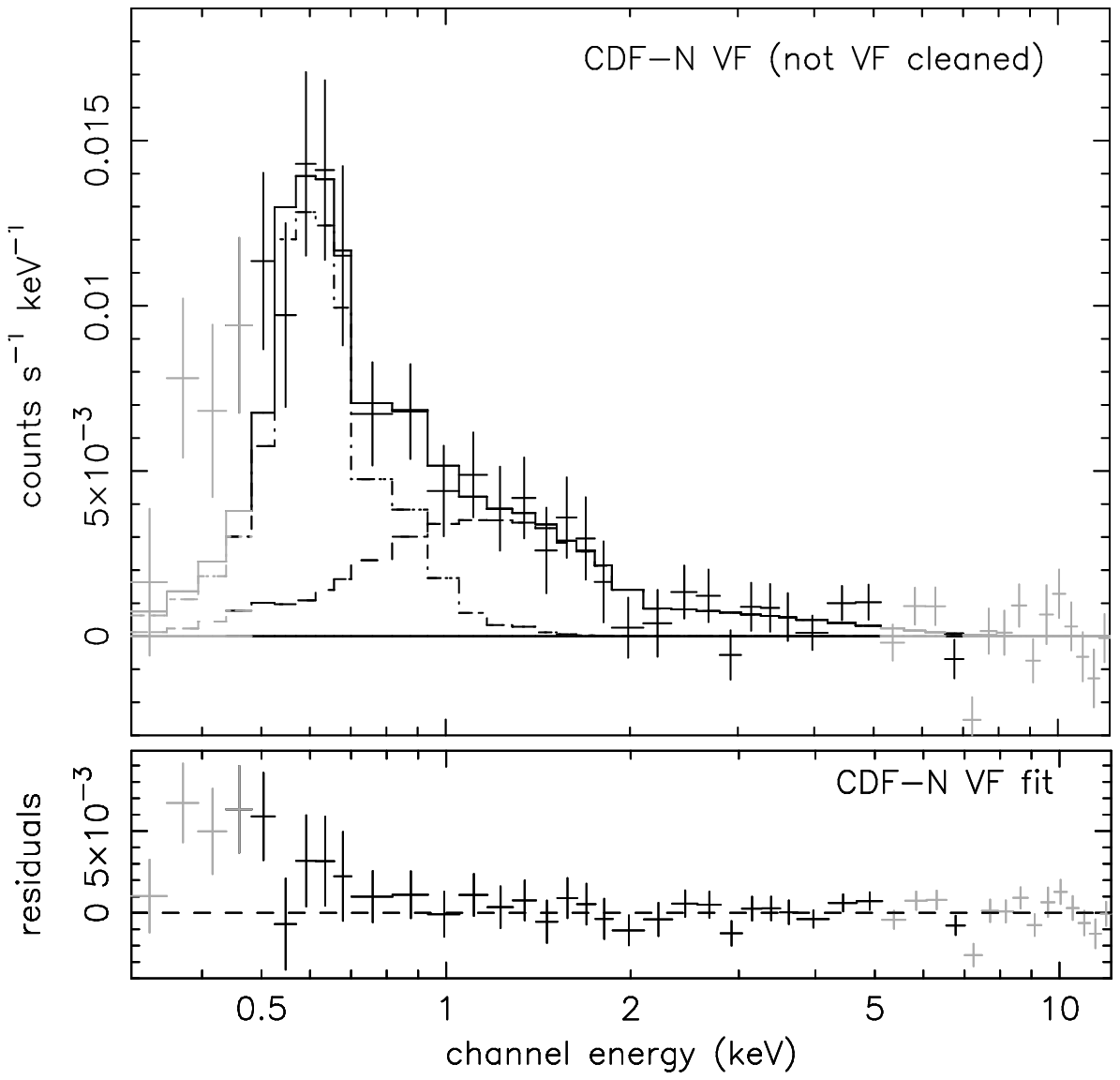}
\caption{Same as Fig.\ \ref{figspec}, but for the CDF-N VF data with
  no VF cleaning. The best fit model to the CDF-N VF spectrum and its residuals are shown,
  indicating a good fit for $E>0.7$ keV. \label{figspecvf_f}}
\end{figure}

\begin{figure*}
\epsscale{1.1}
\plottwo{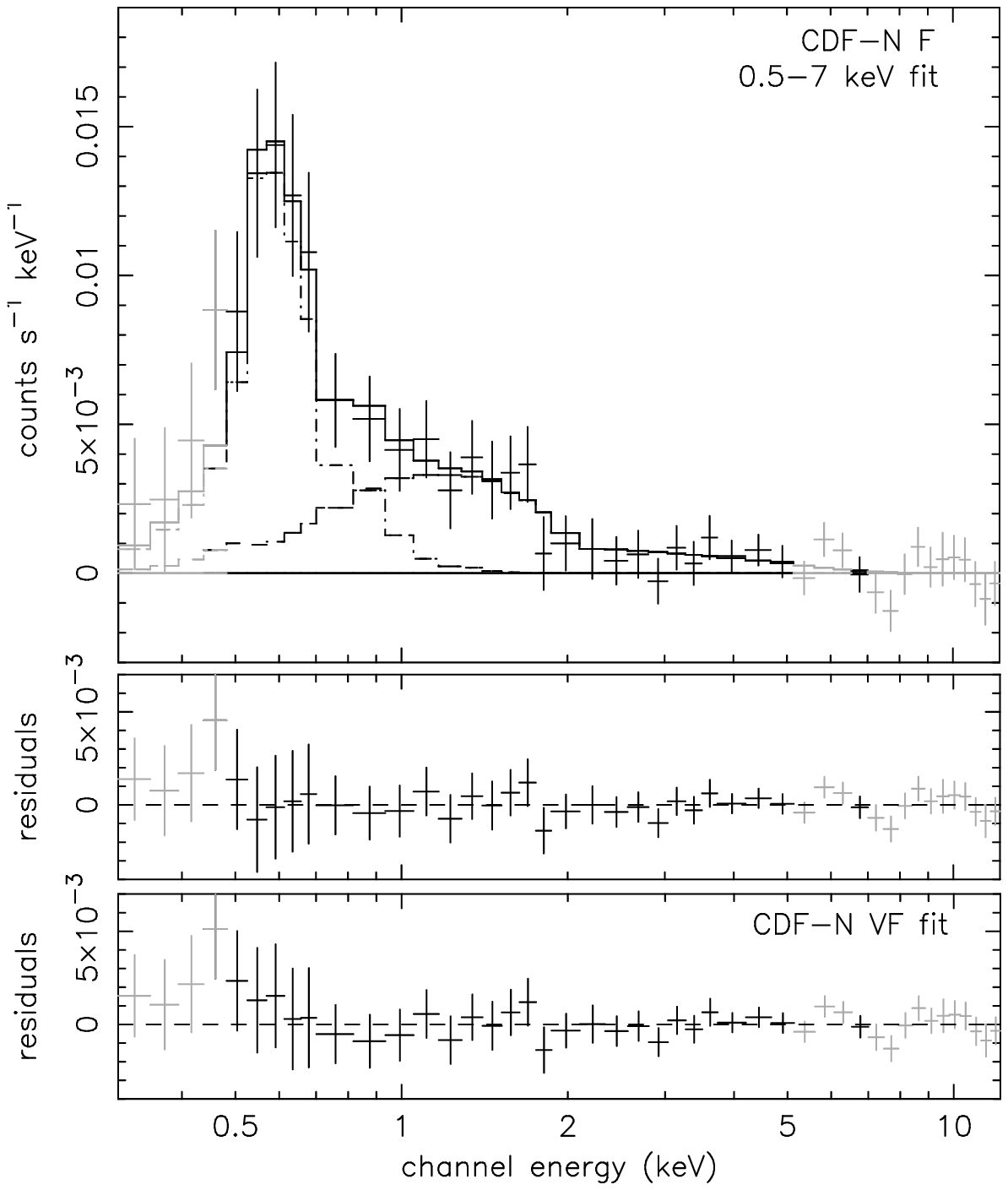}{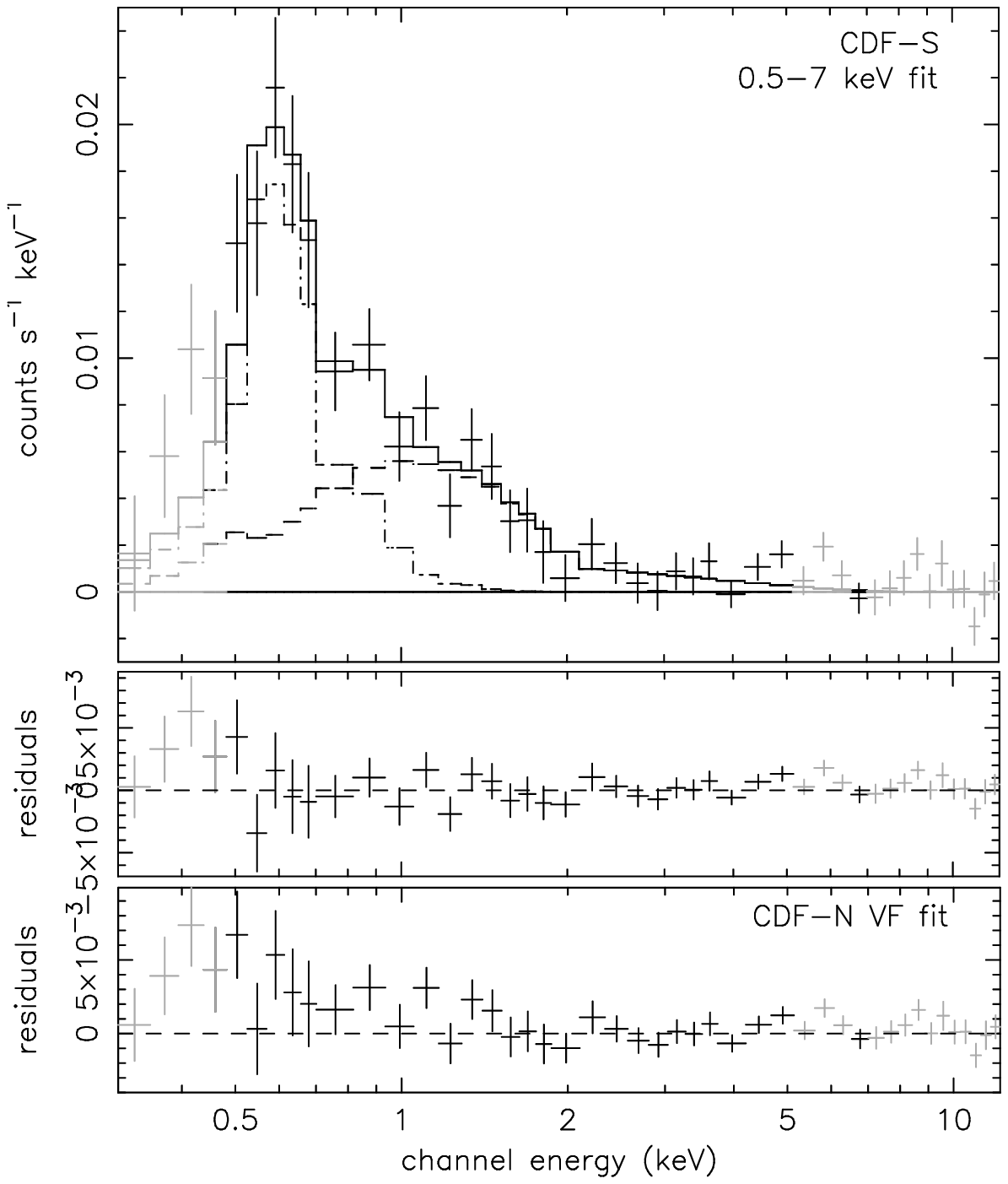}
\caption{Same as Fig.\ \ref{figspec}, but for F mode composite spectra
  for (a) CDF-N and (b) CDF-S.   Fit results are given in
  Table \ref{tblfits}.  Residuals from the best-fit CDF-N VF model are
  shown at bottom. \label{figspecf}}
\end{figure*}

Composite spectra of the unresolved CXB are shown in Figs.
\ref{figspec}--\ref{figspecf}.  There is significant flux below $\sim$2 keV,
consisting of two components: the diffuse soft background, likely a
combination of the Galactic Local Bubble emission \citep{snow04} and charge
exchange emission in regions local to the Sun \citep[e.g.,][]{crav00, warg04}, as well as residual flux from unresolved X-ray point sources
and possibly other CXB components.  The soft diffuse flux consists mainly of
oxygen lines and can be modeled as a thin-thermal plasma \citep[the APEC
model][]{smit01} with $kT\sim0.15$ keV, solar abundances, and zero
interstellar absorption, as in other regions of the sky away from Galactic
features \citep[e.g.,][]{mark03}.  The remaining flux can be modeled as a
power law with a Galactic absorbing column \citep[1.5 and $0.9\times10^{20}$
\cdens\ for CDF-N and CDF-S, respectively,][]{dick90}.

We first fit the spectra in the range 0.5--7.3 keV.  The upper cutoff
here is to avoid bright background lines which appear at $E>7.3$ keV
(Fig.\ \ref{figflare}).  We also exclude bins between 5.6--6.2 keV, to
avoid a faint Mn K$\alpha$ line that apparently makes its way from the
calibration source inside the detector housing.  Because this
radioactive source decays with time, and because the line brightness
at the ACIS-stowed position may be different from that in the normal
position, this line may be subtracted incorrectly when using the
stowed background.  Most other bright background lines are above our
7.3 keV upper cutoff.  For the flux calculations, we also exclude the
2.0--2.3 keV interval to avoid a bright Au line in the detector
background; although it does not affect the fits,
it does add to errors when the background normalization is varied.
Hereafter the spectral analysis and count rates exclude these line
intervals, although the model fluxes are calculated including them.

Fitting results are given in Table \ref{tblfits}.  For our best
dataset, the CDF-N VF spectrum, a combination of the thermal model
plus power law produces a good fit in the 0.5--7.3 keV band, giving $kT_{\rm
APEC}=0.18$ keV and $\Gamma=1.5$.  The thermal component dominates at
$E\lesssim0.8$ keV.  There is some excess above this model for $E
<0.5$ keV (outside our fitting range), which is not surprising because
a single-component APEC model may not be a complete, nor even
physical, characterization of the diffuse soft background
\citep[e.g.,][]{mark03, warg04}.  However the APEC model is sufficient
for our purposes, and gives a very small contribution at $E>1$ keV.

To check that we can compare our fit results for VF-cleaned data to F mode
data, we created a composite spectrum, as above, for the CDF-N VF
subset, but without the VF cleaning.  This spectrum gives very similar
spectral parameters to the VF-cleaned data.  The spectrum is shown in
Fig.\ \ref{figspecvf_f} with residuals from the best-fit model for the
VF-cleaned spectrum.  There is only a small excess above the model for
$E=0.5$--0.7 keV (where VF cleaning removes a large fraction of
background events), but above 1 keV the results are essentially identical.

The composite CDF-N F  and CDF-S spectra give best-fit
spectral parameters that are consistent with the CDF-N VF results
(Table \ref{tblfits}), even though the CDF-S field is half the sky
away and was taken one year earlier.  The spectra are shown in Fig.\
\ref{figspecf}, along with residuals to the best-fit CDF-N VF model.  The CDF-S
spectrum has a somewhat steeper best-fit power law slope
($\Gamma=1.8$) and correspondingly higher normalization, both within
1--2$\sigma$ of the CDF-N VF fits. It is
well-fit by a $\Gamma=1.5$ power law above 2 keV.

\renewcommand{\arraystretch}{1.3}
\begin{deluxetable}{lccc}
\tablecaption{Fits to composite spectra\label{tblfits}}
\tablewidth{0pt}
\tablehead{\colhead{Parameter} &
\colhead{CDF-N VF} &
\colhead{CDF-N F} &
\colhead{CDF-S}}
\startdata
$N_{H}$ ($10^{20}$ \cdens) & 1.5\tnm{a} & 1.5\tnm{a} & 0.9 \\
\cutinhead{Background (0.5--7.3 keV fit)}
power law $\Gamma$  & $1.48_{-0.38}^{+0.46}$ & $1.41_{-0.37}^{+0.43}$ & $1.84_{-0.35}^{+0.48}$\\
$I_{\rm PL}$\tnm{b} & $2.3\pm0.5$  & $2.3_{-0.5}^{+0.5}$ & $3.3\pm0.6$ \\
$kT_{\rm APEC}$ (keV) & $0.177_{-0.016}^{+0.010}$ &
$0.162_{-0.017}^{+0.015}$ & $0.175_{-0.019}^{+0.010}$\\
$I_{\rm APEC}$\tnm{c} & $3.4_{-0.4}^{+0.7}$ & $3.7_{-0.7}^{+1.2}$ &$4.0_{-0.5}^{+1.0}$ \\
\cutinhead{Point sources\tnm{d}}
power law $\Gamma$ & \multicolumn{2}{c}{$1.39 \pm 0.03$} & $1.47\pm0.04$ \\
$I_{\rm PL}$\tnm{b} & \multicolumn{2}{c}{$7.00\pm0.15$} & $6.78\pm0.18$ \\
\enddata

\tnt{a}{\citet{baue04} use a slightly different $N_{H}$ of $(1.3\pm0.4)\times10^{20}$ for CDF-N 
  using values from \citet{lock04} and \citet{star92}.  The difference
  has negligible effect on our results.}
\tnt{b}{Units are photons s$^{-1}$ keV$^{-1}$ ster$^{-1}$ at 1 keV.}
\tnt{c}{Intensities are unabsorbed, in $10^{-12}$ \intens.}
\tnt{d}{Includes point sources in the \citetalias{alex03} catalog,
  within the central 5\arcmin\ of any individual pointing.
We fit CDF-N VF and F as a single dataset.}
\end{deluxetable}
\renewcommand{\arraystretch}{1.2}

For each composite spectrum, we calculate the total observed flux in
the 0.5--1 keV and 0.5--2 keV bands.  For 0.5--1 keV, the power law
contributes 15\% of the flux for both CDF-N spectra, and 20\% for
CDF-S.  We also calculate unabsorbed fluxes in the
extragalactic (power law) component for 1--2 keV (with its
extrapolation to 0.5--2 keV) and 2--8 keV,
performing separate fits for each energy band to minimize their model dependencies.  Because of the limited number of
counts, the fits in these bands do not tightly constrain the power law
slope, which strongly affects the output flux in energy units in the
2--8 keV band.  Therefore, we fix $\Gamma$ and vary only the
normalization; from each best fit model we calculate the observed
flux.  For simplicity, we use values of $\Gamma$ over the 1$\sigma$
errors of the CDF-N VF 0.5--7.3 keV fit (1.1, 1.5, 2.0).  For our final
results we will use $\Gamma=1.5$, and will fold into the uncertainty the
variations in calculated flux between $\Gamma=1.1$ and 2.  We note
that although the CDF-S spectrum has a best-fit $\Gamma=1.8$, for
0.5--7.3 keV, we do not introduce a significant error into the flux values
because we fit the spectrum normalization separately in the individual
bands.  For all wide-band flux measurements in this paper, we include an ACIS
flux calibration uncertainty of 3\%\footnotemark.

\footnotetext{see {\tt http://cxc.harvard.edu/cal/}.}

\subsection{PSF scattering}\label{psfscatter}
While we have used large exclusion regions to remove the contribution
of point sources, a small fraction of the source counts, in the wings
of the PSF, will not be excluded and must be subtracted from our
unresolved fluxes.  To estimate this scattered flux we use a model
(from the \chandra\ CALDB) which gives the shape of the ACIS PSF as a
function of energy and off-axis angle.  Using the \citetalias{alex03}
catalogs for CDF-N and CDF-S sources, we calculate the PSF for each
source inside the 5\arcmin\ extraction radii.  We thus determine the
total count rate for scattered source photons that are outside the
exclusion regions.  We find that in the 1--2 keV band, 1\% (CDF-N) and
0.5\% (CDF-S) of the unresolved fluxes come from scattered source
photons.  For 2--8 keV, where the PSF is broader, this fraction is 4\%
for CDF-N and 2\% for CDF-S.  Hereafter, unresolved fluxes have been
corrected by these factors.

\renewcommand{\arraystretch}{1.1}
\begin{deluxetable*}{lcccc}
\tablewidth{5in}
\tablecaption{Unresolved fluxes from composite spectral fits\label{tblflux}}
\tablehead{
\colhead{} &
\colhead{CDF-N VF} &
\colhead{CDF-N F} &
\colhead{CDF-S} &
\colhead{Average\tnm{a}}}
\startdata
\cutinhead{Observed (power law + APEC)}
0.5--1 keV & $4.0\pm0.3$ & $4.3\pm0.5$ & $5.0\pm0.4$ & \nodata \\
0.5--2 keV & $4.9\pm0.6$  & $ 5.3\pm 0.6$ & $6.0\pm0.5 $ & \nodata \\
\cutinhead{Extragalactic (unabsorbed power law)\tnm{b}}
0.5--2 keV\tnm{c} & $ 1.58\pm 0.34$ & $ 1.68\pm 0.39$  & $ 2.05\pm
0.41$ & $1.77 \pm0.31$\\
1--2 keV& $ 0.93\pm 0.17$ & $ 0.99\pm 0.20$  & $ 1.20\pm 0.19$ & $1.04\pm0.14$ \\
2--8 keV & $ 3.5\pm 2.4$   & $ 3.1\pm 2.4$    & $ 3.6\pm 2.2$ & $3.4\pm1.7$\\
\enddata
\tablecomments{Intensites are in units of $10^{-12}$ \intens.}
\tnt{a}{Intensities below 1 keV are not expected to be the same in different
  fields, so they are not averaged.}
\tnt{b}{Power law intensities are for $\Gamma=1.5$.  Detailed error
analysis for the 1--2 keV and 2--8 keV values are given in Table \ref{tblerr}.}
\tnt{c}{0.5--2 keV intensities are extrapolated directly from the 1--2
  keV fits.}

\end{deluxetable*}

\subsection{Calculation of background intensity}\label{calcdiff}
To convert the unresolved flux into intensity, or sky surface
brightness, we divide the observed flux by the effective solid angle
subtended on the sky by the extraction region.  To find this solid
angle, we create an exposure map for the \mbox{ACIS-I} array for each
observation, normalized to a maximum value of 1.  We do not include
the effects of the mirror vignetting and CCD efficiences, as these are
included in the ARF during the flux calculations, so that only the
chip coverage, the excluded columns on the CCDs, and the telescope
dither are included in the image.  We then integrate this image within
the sky spectrum extraction regions (described in \S\ \ref{specextr}),
to give the effective solid angle for the unresolved spectrum.  For
the composite spectra, we average these solid angles weighted by the
exposure time of each observation, giving 0.0135 deg$^2$ for both
CDF-N VF and F, and 0.0159 deg$^2$ for CDF-S.  These are used to
calculate the unresolved background intensities, which are given in
Table \ref{tblflux}.

For measuring the intensity due to sources (see \S\ \ref{sourspec}), we
perform a similar solid angle calculation, including the chip coverage
and telescope dither, for the entire central 5\arcmin\ region (without
the exclusion regions).  The exposure weighted averages of these solid
angles are 0.0202 deg$^{-2}$ for CDF-N and 0.0200 deg$^{-2}$ for
CDF-S.

\subsection{Average unresolved intensities}\label{calcdiff1}
Under the assumption that the unresolved CXB is isotropic, we
calculate mean values of the unresolved CXB intensity using
measurements from the three data subsets.  We take care to treat the
statistical errors in the stowed background dataset (which comprise
$\sim$40\% of the total error) as not independent between the subsets.
We therefore average only those errors that arise from statistics in
the observed sky count rates, and in the 2\% uncertainty in the
detector background shape (which should be independent for the three
subsets, which are separated in time by $\sim$1 year).  Mean values
for the unresolved intensities are given in Table \ref{tblflux}.
Details of the error propagation is shown in Table \ref{tblerr}, which
is described in detail in \S\ \ref{sumerr}.

\section{Verification}\label{verif}

In this section we test the dependence of our measurement on the
details of the source exclusion and background modeling,
and also consider the possibility of residual background flares
significantly contaminating the unresolved signal.

\subsection{Contribution from fainter sources}\label{faintsource}
The source catalog from \citetalias{alex03} that we use here was
derived with completeness in mind, and so was limited to sources
brighter than a certain flux.  We test here whether fainter but still
detectable sources contribute significantly to the observed signal.
To detect sources at lower significance than \citetalias{alex03}, we
use all the CDF exposures \citepalias[including those before 2000
  January 21, see][Table 1 and A1 for full lists]{alex03} to
create a deep image for each field (Fig.\ \ref{figim}).  After a less
conservative cleaning for bright flares, these images have exposure
times of 1.9 Ms for CDF-N and 0.8 Ms for CDF-S.  We detect sources in
three bands, 0.5--2, 2--8, and 0.5--8 keV, corresponding to the
``soft'', ``hard'', and ``full'' bands of \citetalias{alex03}.  We run
the wavelet decomposition \citep{vikh98} on these deep exposures, and
measure centroids and fluxes using the largest-scale wavelet
decomposition as the background.  For source exclusion purposes, it is
not necessary to match the sources detected in different bands, even
though the source exclusion regions will overlap.  Our detected source
list therefore consists of a concatenation of the sources detected in
the individual bands.

\begin{figure}
\epsscale{1.2}
\plotone{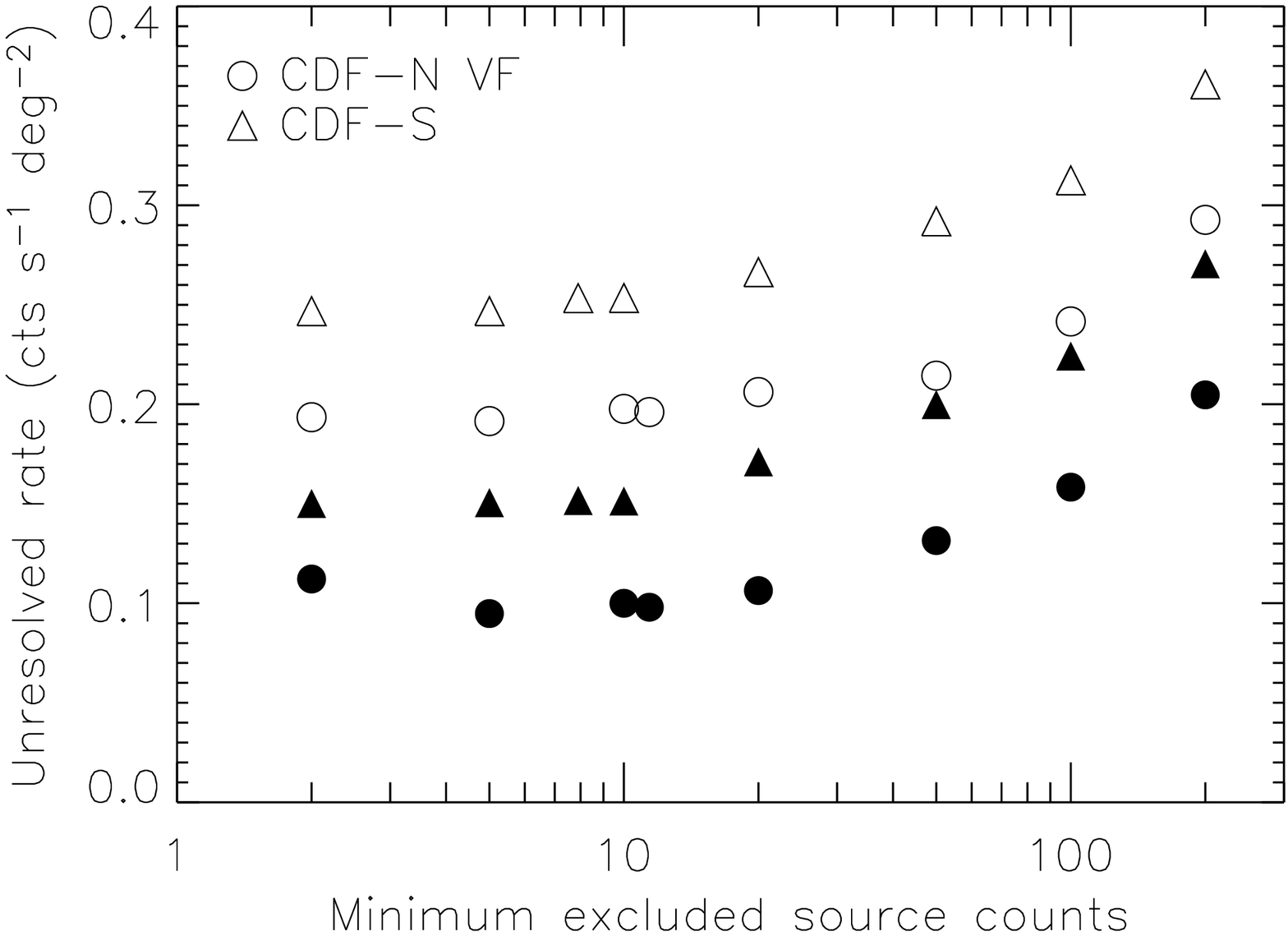}
\caption{Unresolved CXB intensity, excluding sources
  with different minimum numbers of counts in the 0.5--8 keV band.
  Open symbols are for 1--2 keV, filled symbols are for 2.3--7.3 keV.
  Errors are not shown for clarity. The differences between the fields
  are within their statistical uncertainties. Note that the slight upturn at
  2 counts per source for CDF-N VF 2.3--7.3 keV is not significant.
  The increase in intensity is due to the smaller solid angle
  over which the unresolved flux was calculated.  \label{figflux}}
\end{figure}

We sort the detected point sources by the number of photons associated
with each source (total photons minus background) for the 0.5--8 keV
band.  We then extract the unresolved spectrum (for CDF-N VF and CDF-S), after excluding
sources with different minimum numbers of photons.  In the
\citetalias{alex03} catalog, the minimum number of detected photons was
11.4 for CDF-N and 8.0 for CDF-S.  However for our test we have
detected sources at very low significance down to 2 photons above the
background.  By excluding the contributions of these possible sources
as well, we can determine if such low-flux objects contribute
significantly to the total flux.

In Fig.\ \ref{figflux} we show the unresolved intensity, as a function
of the minimum number of counts from excluded sources.  It is apparent
that sources with 2--10 source photons do not contribute significantly
to the background flux.  We note that while many of these fainter
``sources'' are not real and are only due to statistical fluctuations,
any real sources are detected and excluded.  This confirms that our
results are insensitive to the details of source detection. 

We also have directly checked that we are fully excluding scattered flux from
sources, by re-calculating the unresolved CXB intensities using exclusion
radii that are 30\% larger for point sources, and twice as large for
extended sources.  We find that the results agree to $\lesssim$3\%, as
expected (\S\ \ref{psfscatter}).

\subsection{Variations in I1 chip}\label{vari1}
Because the stowed observations did not include the I1 chip (which is
modeled using the events from the I0 chip), it is
possible that the stowed background spectrum is somewhat
different for this chip and thus is not properly subtracted in our
analysis.  We have tested this by performing the spectral analysis including
only chips I0, I2, and I3, and find no changes in the unresolved intensities and spectral
parameters to $<$2\%.  Therefore the I1 chip background subtraction
has negligible effect on the results.

\subsection{Contributions from flares?}
Because of the very small unresolved signal, even a very low-level
residual background flare contribution can affect our results.  The 2--8 keV count rate for the instrumental
background is a factor of $\sim$25 larger than that for our unresolved
sky signal, so even a flare on the level of a few percent can have a significant
effect.  Therefore it is important to rule out the
existence of such low-level flares.

\begin{figure*}[t]
\epsscale{1.}
\plotone{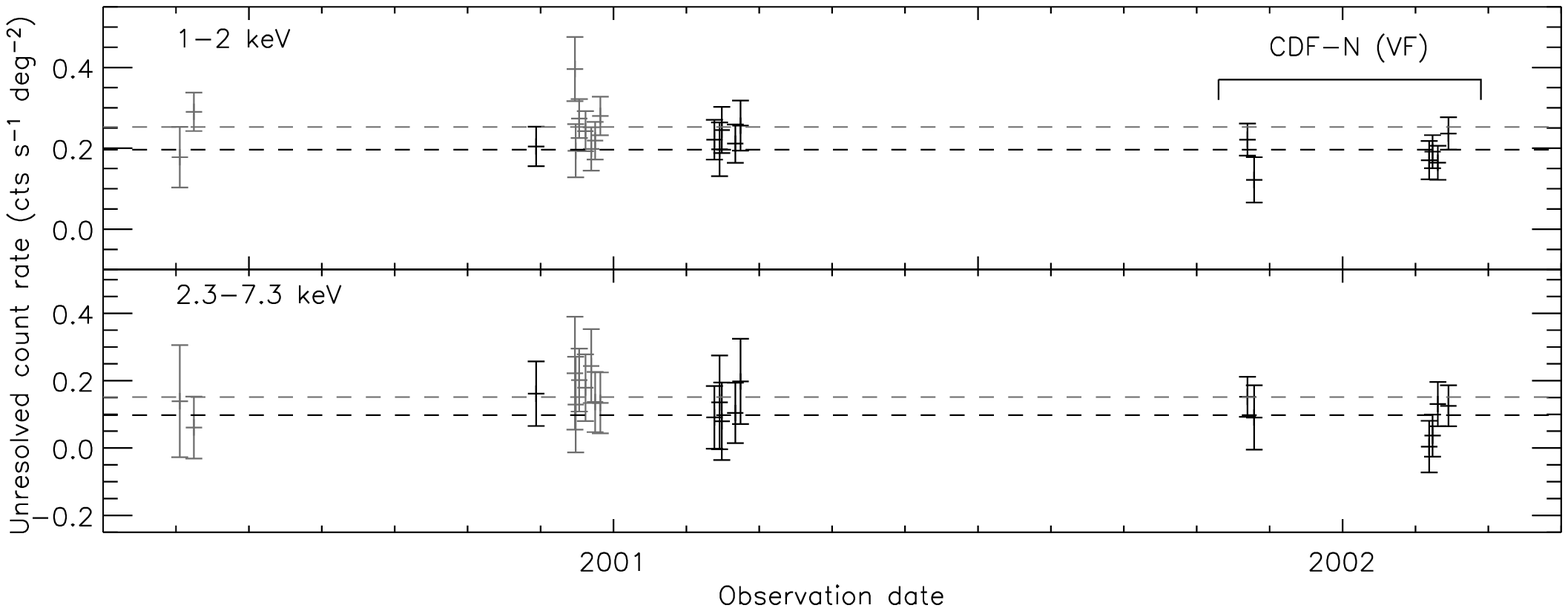}
\caption{Points with error bars show the unresolved CXB intensity in
  the range 1--2 keV (top) and 2.3--7.3 keV (bottom), for 21 CDF
  observations that are included in the spectral analysis (CDF-N in black, CDF-S in gray).  Shown with dashed lines is the count
  rate for the composite CDF-N VF and CDF-S observations.   \label{figtime}}
\end{figure*}

We first test for flares by directly examining the observed
unresolved spectrum.  Fig.\ \ref{figflare} shows a sum of several
observed flares in \mbox{ACIS-I} 
spanning a number of years and an order of magnitude in
intensity, giving a representative average flare spectrum
that we could possibly expect in our long datasets. A
spectrum from the quiescent periods was subtracted for each
included observation, so only the flare excess is shown.
The individual flares have similar (although not identical)
general spectral shape, regardless of the brightness; the
relative intensity of the fluorescent lines, if present,
varies from flare to flare (always being lower than in the
quiescent spectrum). The flare spectrum in Fig.\ \ref{figflare} is
normalized to have the same 9--12 keV rate as the quiescent
spectrum.  Clearly, it has a shape that is very different from the
quiescent background, and from the shape of the much more
frequent BI-only flares \citep{mark03}. What is
most useful for us is that the flares are also very different from
the X-ray sky signal, increasing steeply towards low
energies where the sky spectrum is absorbed by the CCD
filter. We can therefore use spectra at low energies to put
constraints on flare contamination.
                                                                                
When the normalized quiescent background is subtracted from the flare
spectrum in Fig.\ \ref{figflare} (as it would be in our analysis if it were present in a sky
spectrum), its spectrum in the 0.3--7 keV band, excluding the 1.8--2.2
keV line interval, can be described (without the application of an
ARF) as a sum of two power laws, with soft $\Gamma=1.7$ and hard
$\Gamma=-0.1$, with a 13\% contribution of the soft component to the
total 0.3--7 keV count rate.

We fit the unresolved spectra, including such a flare model, in the
0.3--7 keV band.  We add the two power law model above, fitted with no
ARF, to the same APEC plus power law spectrum as in \S\
\ref{specanal}.  For all three composite spectra, we find that the
flare component can contribute at most $\sim$14\% of the flux
(1$\sigma$ upper limit) in the 1--2 keV band and $\sim$35\% in the
2--8 keV band.  The addition of this flare model does not
significantly improve the reduced $\chi^{2}$ of the fit.
                                               
We should note that a different, very rare FI flare species
was observed, consisting mostly of the high-energy
fluorescent lines\footnote{\raggedright \chandra\ 2003 Calibration
Workshop,
{\tt http://cxc.harvard.edu/ccw/proceedings/03\_proc/ 
presentations/markevitch2}},
but it could not significantly affect our results due to its
spectral shape.

We can also test for flares by noting that flare activity is generally
time-variable and so might be expected to give a strong variation in
the background-subtracted count rates between observations.  To check
this, we subtract background as described in \S\ \ref{instback} for
each of the individual exposures, and calculate a count rate scaled by
the effective solid angle on the sky (see \S\ \ref{calcdiff}), for the
1--2 keV and 2.3--7.3 keV bands, shown in Fig.\ \ref{figtime}.  Errors
include the statistical count rate uncertainty, as well as the
statistical uncertainty in the background normalization (but not its
systematic uncertainty).  In both
bands, the count rates over time are consistent with those from the
CDF-N and CDF-S composite spectra.  Note that because we normalize the
quiescent background using the 9--12 keV count rate, our filtering
using the ratio of the 2.3--7.3 keV and 9--12 keV count rates (\S\
\ref{flarback}) should make, by design, the residual 2.3--7.3 count rates equal to within the
statistical errors.  The larger statistical scatter is due to the fact
that the ratio filtering was performed on the full \mbox{ACIS-I} field,
whereas here the fluxes are from the $r<5$\arcmin\ region, which is 5
times smaller.  We stress, however, that we have not performed any
light curve filtering using the 1--2 keV band, so the fact that the
1--2 keV signal is not dominated by a few bright observations implies
that there is no significant flare contribution.

Thus, any flare contribution could only be due to some low-level,
persistent background flare that is roughly constant over many
observations spanning several years.  One could possibly argue that
our filtering selects a low-end tail of the flare distribution, but this is
not the case.  Recall that 20 ks bins with
spectral deviations outside 3\% from the global average were excluded
during the temporal cleaning (\S\ \ref{flarback}).  This filtering
excludes 24\% of the total exposure, but most of these excluded
deviations are quite small; in fact, only two (for ObsIDs 3294 and
3390) appear to be inconsistent with purely statistical scatter.
(Fig.\ \ref{figrat}).  We note as well that the blank sky observations
shown in Fig.\ \ref{figblank} have not been filtered using the band
ratio as we did here, and they still show a rather constant flux in
the 2--7 keV band over several years.   We conclude that any significant flare
contamination is very unlikely.  We also remind here that there is no
constant background component that is present in the sky data but absent in the stowed background, as shown by comparison of the stowed
and dark Moon observations (\S\ \ref{blockcomp}).

\section{Discussion}\label{resofrac}

\subsection{Comparison between datsets}
The unresolved CXB intensities (Table \ref{tblflux}) are consistent to
1$\sigma$ for the three datasets; the difference between CDF-N and
CDF-S is $<$25\%.  We note that although the source detection limits
in these two fields are different by a factor of two, the contribution
of sources with fluxes between these limits is relatively small. Given
typical \logn\ distributions \citepalias[see \S\
\ref{sourcedist}]{more03}, sources between the two flux limits should
only contribute $\sim$$6\times10^{-14}$
\intens\ for 1--2 keV and $3\times10^{-13}$ \intens\ for 2--8 keV,
which are small compared to the unresolved fluxes and our measurement
errors.  Although we cannot make definitive statements using only two
fields, this correspondence between CDF-N and CDF-S hints that the
distribution of X-ray sources at fluxes below the current detection
limits (or any diffuse component) is isotropic.  This is especially
interesting given that the CDF-S contains significantly fewer detected
point sources \citep[e.g.,][]{bran01b, rosa02}.

\subsection{Total X-ray background}\label{totalback}

To compare our results to other studies, we determine the total X-ray
background intensity in the ranges 1--2 and 2--8 keV, by adding our
measured signal to flux from detected sources from the CDF
observations and from wider-field surveys.  We can then use the total
CXB value to obtain the resolved fraction of the CXB.  We note,
however, that it is the absolute unresolved flux given above, not the
resolved fraction, that usually matters for source population studies.

\subsubsection{Spectra of the excluded sources}\label{sourspec}

\begin{figure*}
\plottwo{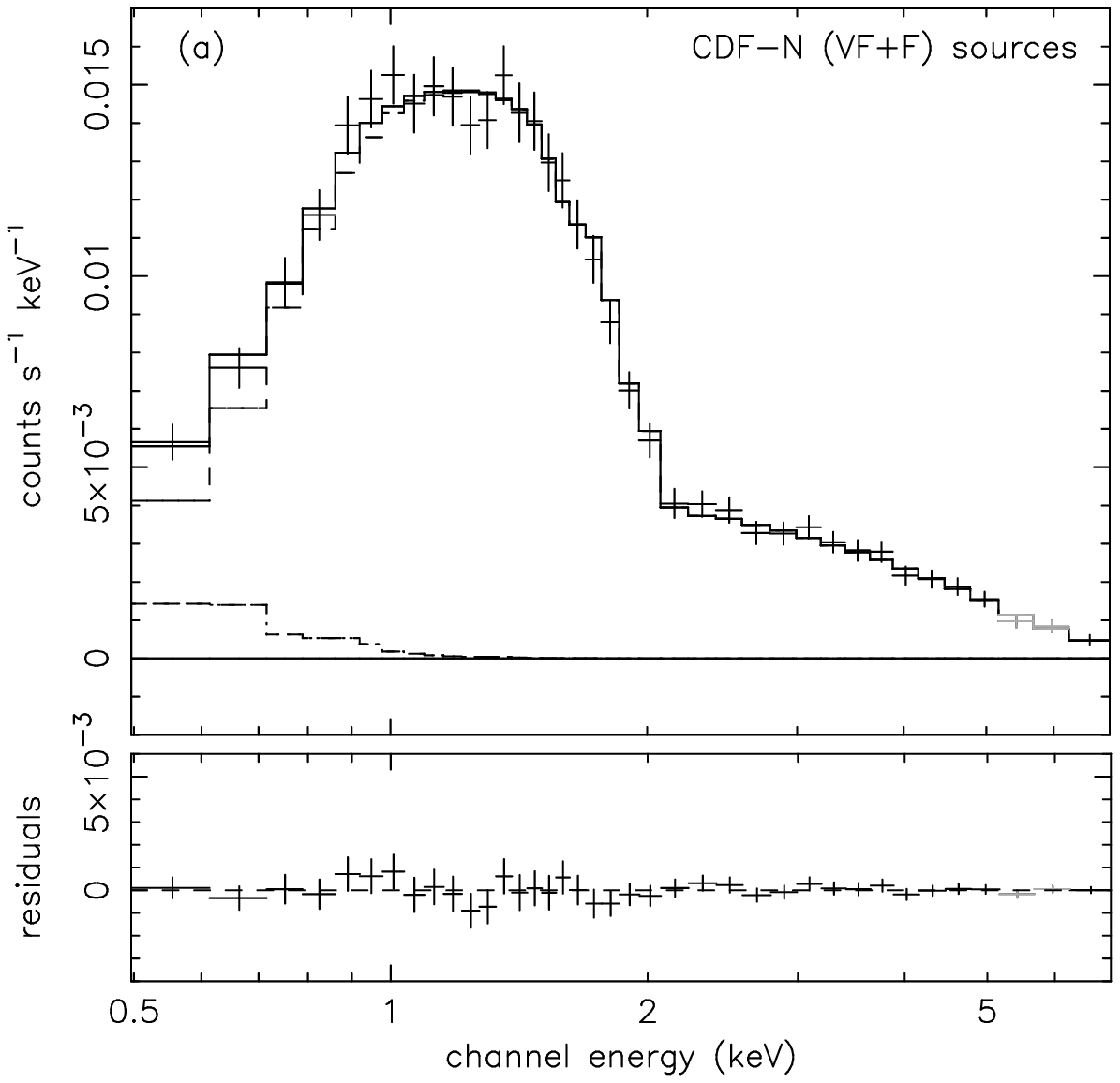}{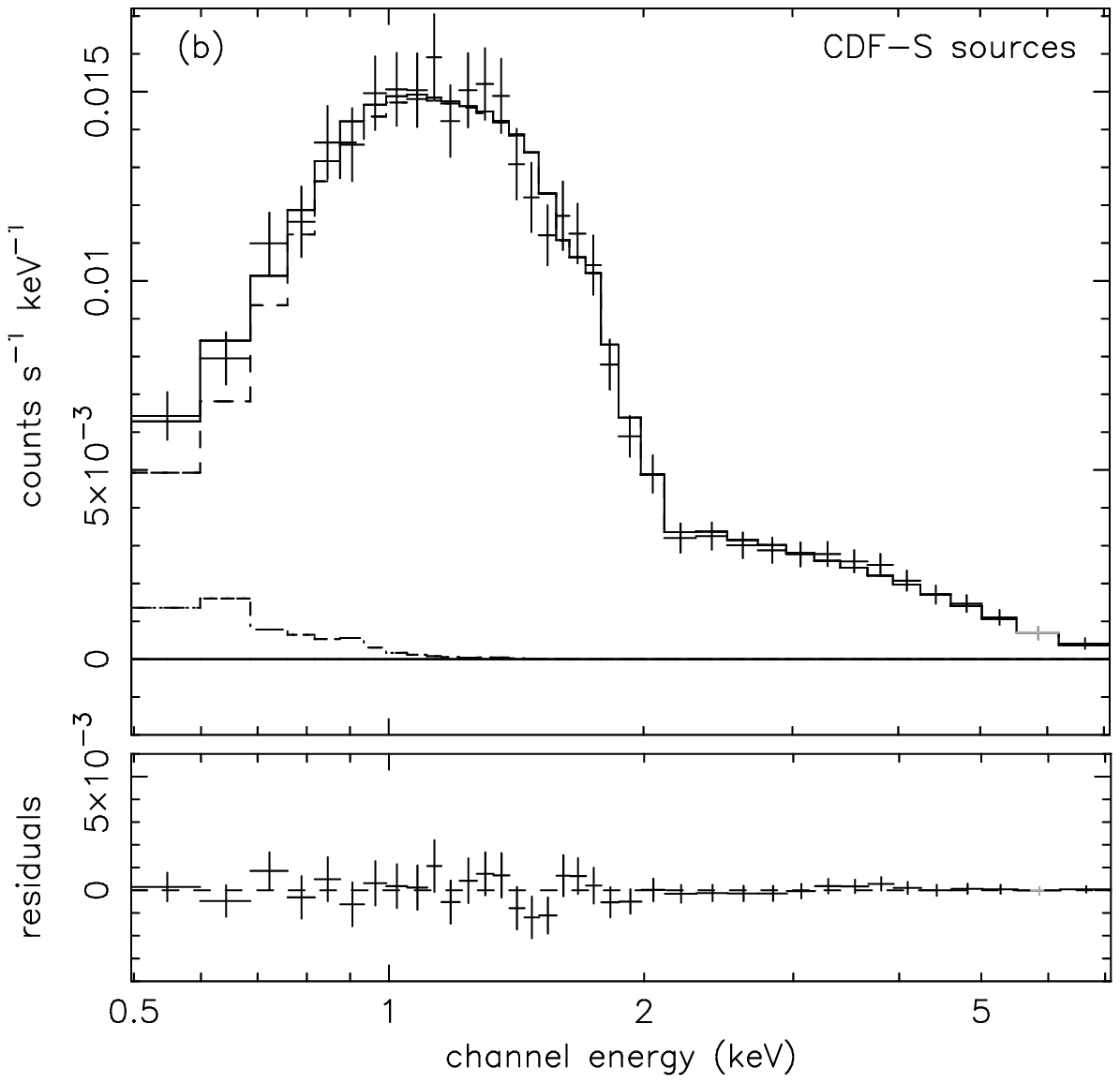}
\caption{Total spectra for detected sources from the
  \citetalias{alex03} catalog contained within 5\arcmin\ radius of the
  aimpoint for any observation, for the (a) combined CDF-N (VF + F) and
  (b) CDF-S datasets.  The separate APEC and power law components are
  shown as dashed lines.  Spectral
  fits are given in Table \ref{tblfits}.\label{figsources}}
\end{figure*}

\begin{figure}
\epsscale{1.25}
\plotone{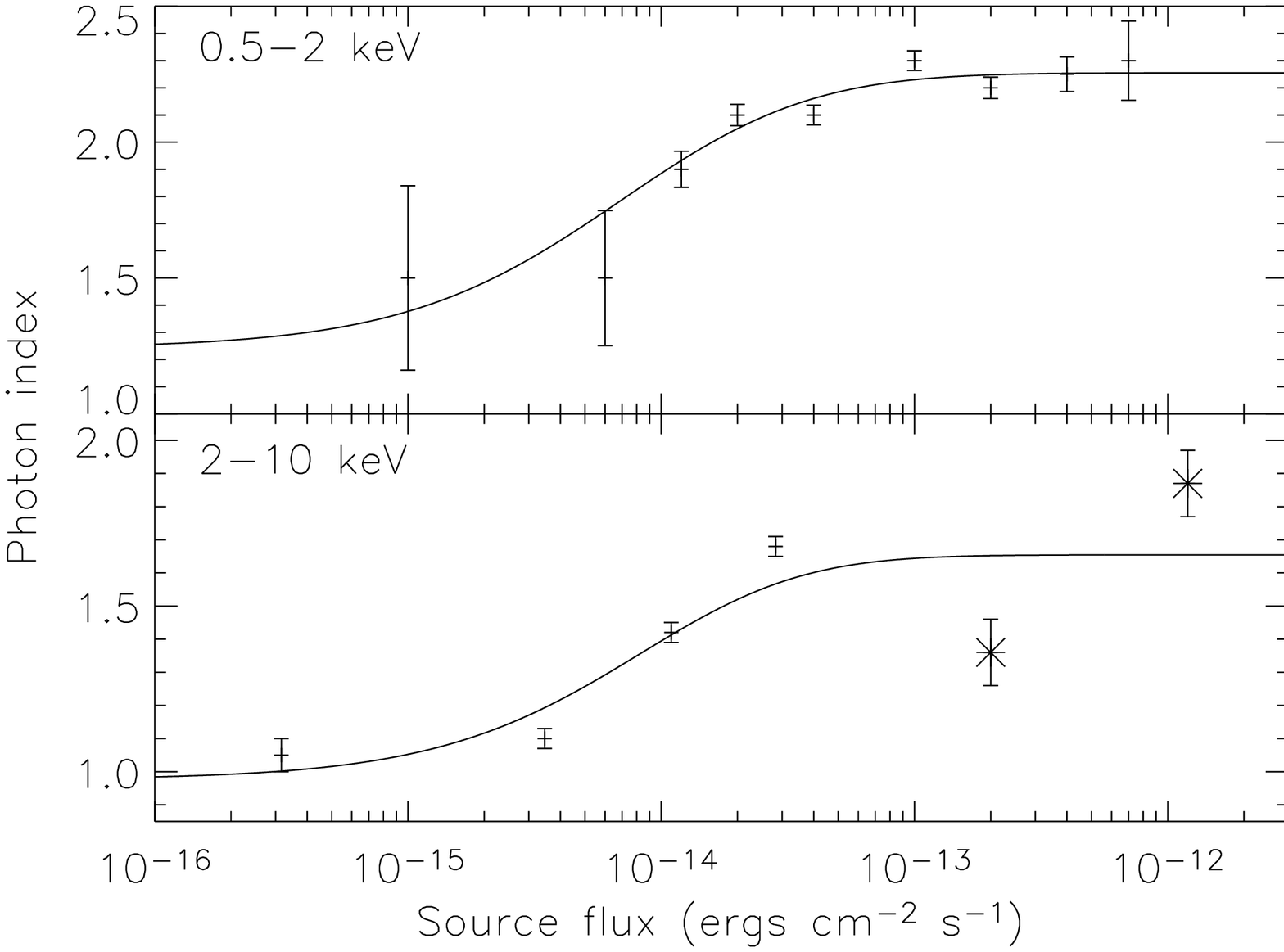}

\caption{Variation in power law photon index $\Gamma$ with source
  flux, for (a) 0.5--2 keV and (b) 2--10 keV sources.  Data for
  0.5--2 keV are from the \rosat\ study of \citet{vikh95a}, for 2--8 keV
  data are from \asca\ \citep[stars][]{dell99} and CDF-S \citep[crosses][]{rosa02}.  In both
  cases the flux dependence of $\Gamma$ is fit with a scaled error
  function (solid lines).  \label{figgam}}
\end{figure}

First, we extract the total spectrum of the detected point sources for
each of the two fields.  To limit the effects of vignetting, for each
individual observation we only include sources from inside the
5\arcmin\ radius.  Because there are variations in the pointing
between the observations (especially for CDF-N), the 5\arcmin\ circle
covers slightly different patches of sky for each observation.  Thus
the total solid angle sampled is 0.0331 deg$^{-2}$ for CDF-N and
0.0237 deg$^{-2}$ for CDF-S.  However, each individual spectrum covers only the
central 5\arcmin\ circle, so when we add the spectra, the
effective sky coverage of the total spectrum is an exposure-weighted
mean of the sky coverages of the individual 5\arcmin\ regions.
Therefore, we use the solid angles as given in \S\ \ref{calcdiff} to
calculate the total source flux deg$^{-2}$.

\begin{figure*}
\epsscale{1.1}
\plottwo{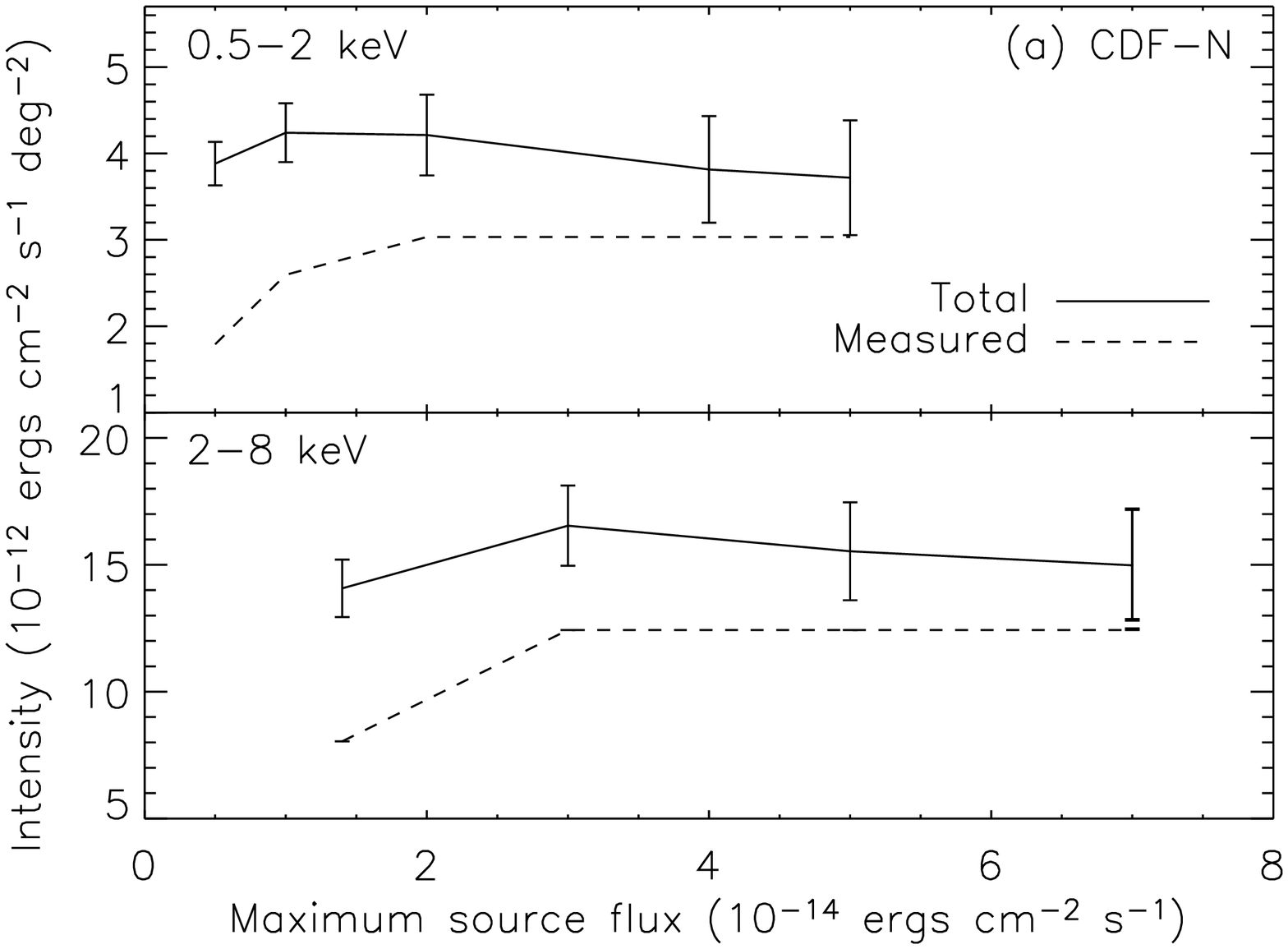}{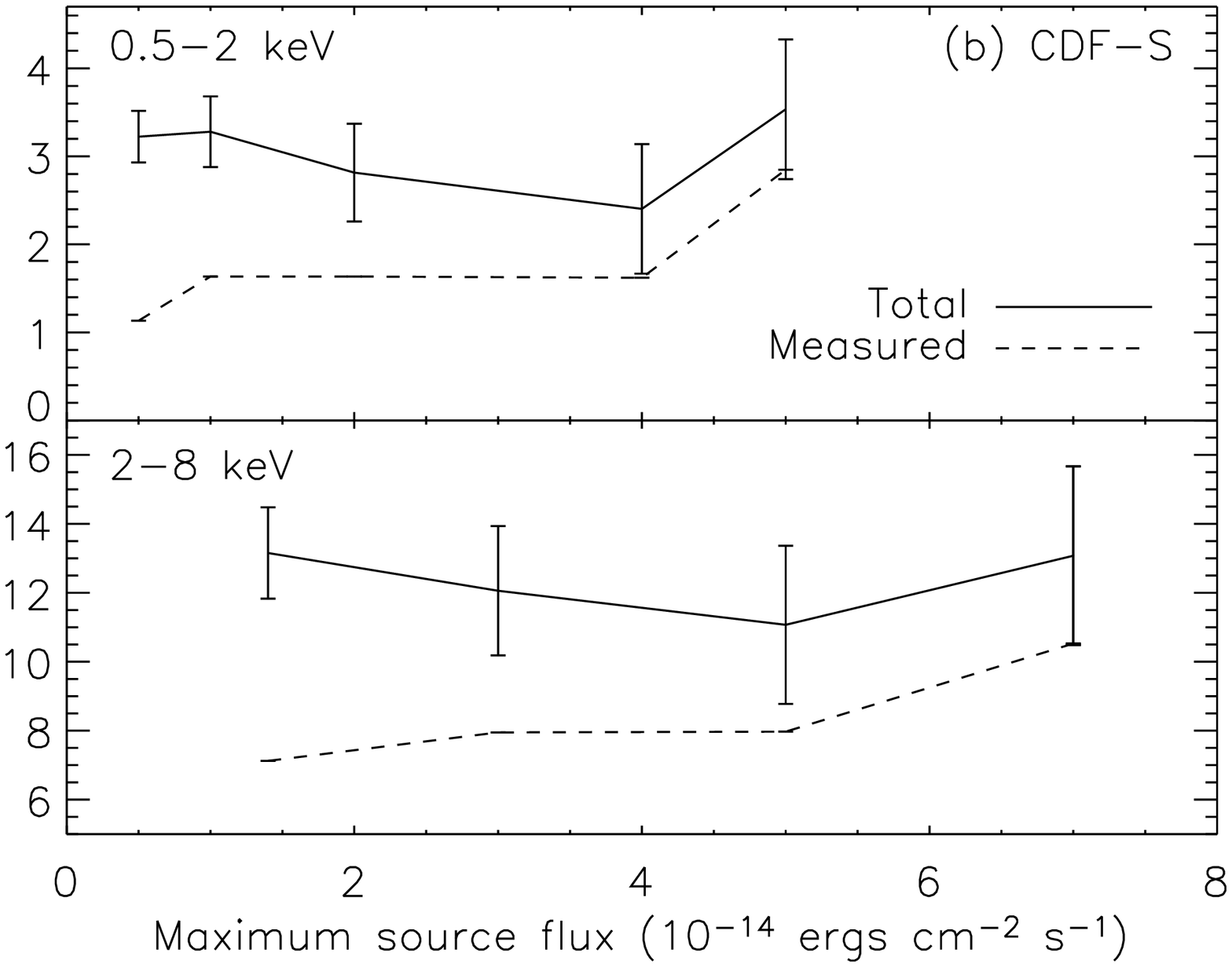}
\caption{Measured and total (including bright source correction)
  fluxes for our composite source spectra, as a function of the upper flux
  cutoff (\S\ \ref{sourcedist}) for (a) CDF-N and (b) CDF-S.  Fluxes for
  sources in the soft band are given for 0.5--2 keV, while total
  intensities are for 1--2 keV.  Errors shown on
  the total fluxes include measurement error and Poisson error in both
  the measured source fluxes and in the bright source correction.
  \label{figcutoff}}
\end{figure*}

\begin{figure*}
\epsscale{1.2}
\plotone{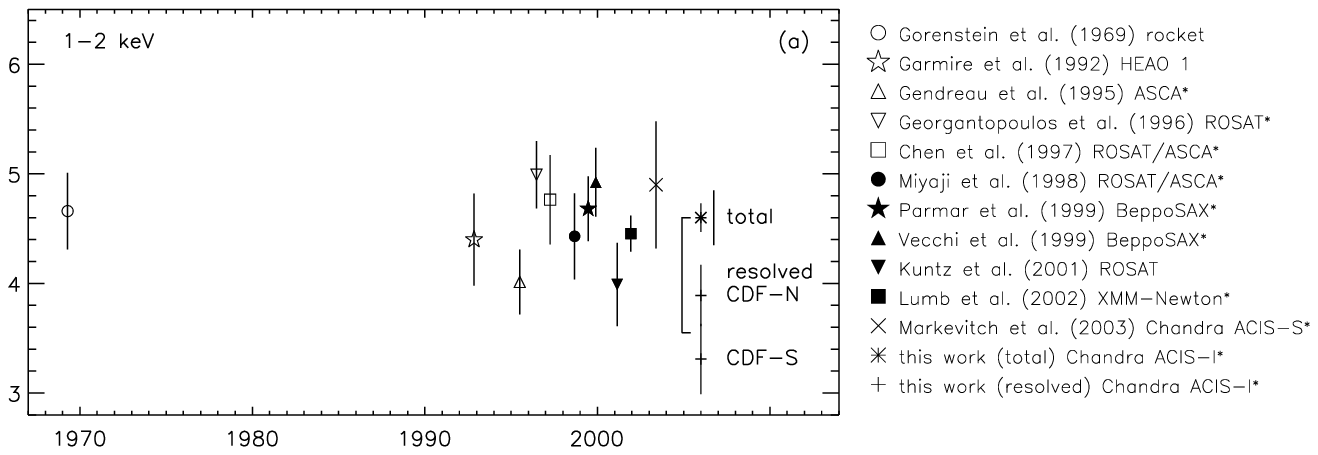}
\vskip-1.25in
\plotone{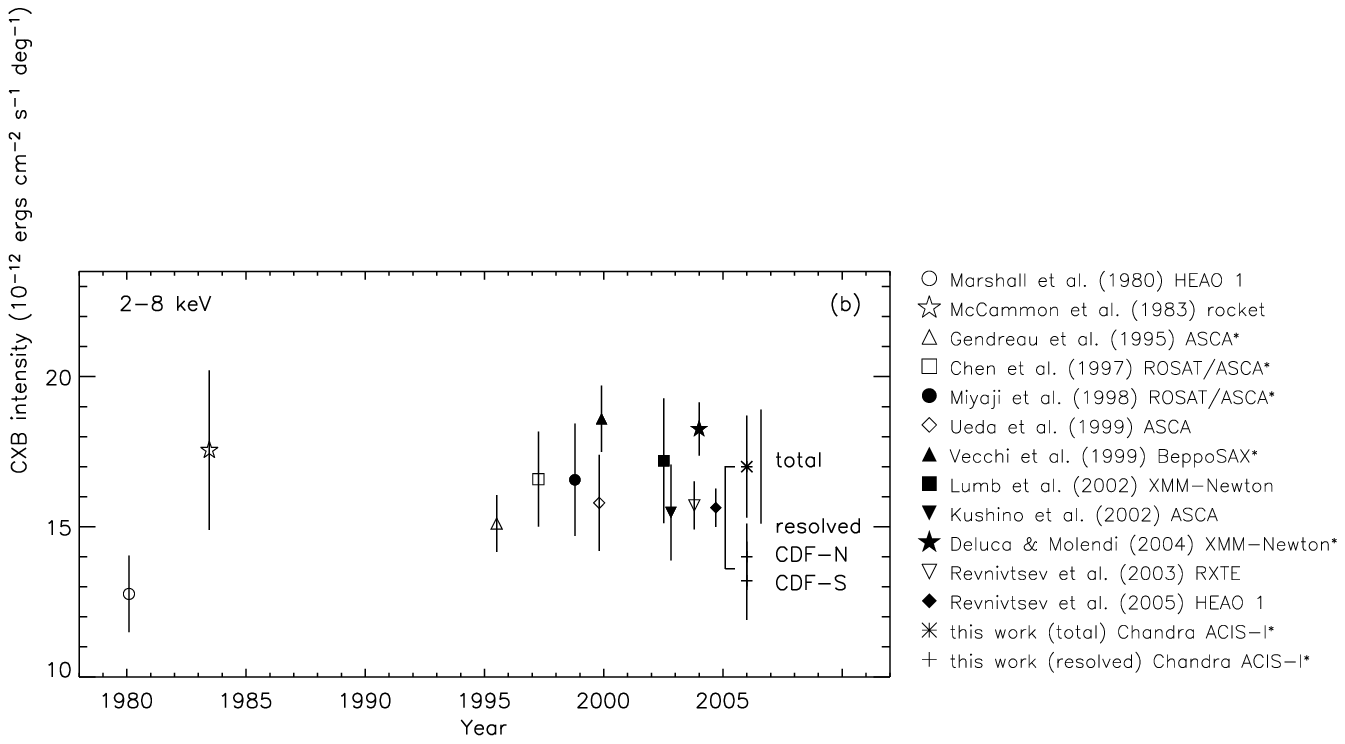}

\caption{Measurements of the total (a) 1--2 keV and (b) 2--8 keV CXB.
  For our results, the relative contributions of sources from CDF-N and
  CDF-S are shown as two separate crosses; the totals are calculated
  from the average of these.  Our measured unresolved intensities are
  shown by brackets.  The small error bar on our total CXB represents
  only the error in the unresolved component.  Full error bars,
  including uncertainty in bright source corrections, are shown just to the right.
  References marked with an asterisk have been corrected for the
  contributions of bright sources (Table \ref{tblcorr}).  Note that
  these values are somewhat different from a similar plot in
  \citetalias{more03}.
  \label{figtot}}
\end{figure*}

We extract the source spectra in regions as described in \S\
\ref{sourexcl}, although with a smaller radius of $1.5r_{90}$ for each
source, so as to include most source flux but to limit the
diffuse background contribution.  Composite spectra for CDF-N
(combining F and VF) and CDF-S are shown in Fig.\ \ref{figsources}.
We fit the spectra with the APEC plus a power law component.  The APEC
model is necessary because the source extraction regions still have
sufficent area to detect the soft diffuse component; fit results are
given in Table \ref{tblfits}. 

To calculate the total source intensities, we must perform
aperture corrections to account for small amounts of scattered flux outside the
extraction region. We use the \citetalias{alex03} catalogs and the
ACIS PSF model as described in \S\ \ref{psfscatter}, and find aperture
corrections of 5\% (CDF-N) and 3\% (CDF-S) for the 1--2 keV band, and
9\% (CDF-N) and 6\% (CDF-S) for the 2--8 keV band.
The aperture-corrected source intensities in two bands are given in Table
\ref{tblsrc}. The brightest sources in these fields have 0.5--8 keV
fluxes of $3.5\times10^{-14}$ and $1.1\times10^{-13}$ \flux\ for CDF-N
and CDF-S respectively, and the lowest fluxes are at the
\citetalias{alex03} source detection limits \citepalias{alex03}.

\subsubsection{Total background using \logn}\label{sourcedist}
We cannot directly use the source fluxes determined above for CDF-N
and CDF-S to determine the total background, because the narrow CDF
fields do not include a representative number of bright, rare sources
that contribute a significant fraction of the total.  We must include
their contribution based on source counts from published wider-area
surveys.  Furthermore, to reduce the effective Poisson scatter of the
number of the rarest sources in our fields, we should remove the
sources brighter than a certain flux \citep{hasi96, lumb02} which we
will determine below.

In the soft band, \citet{vikh95a} obtained a 0.5--2 keV
\logn\ curve from a wide-area (20 deg$^2$) \rosat\ PSPC survey for sources with fluxes between
$1.2\times10^{-15}$ and $7\times10^{-13}$ \flux.  We can directly use
these data to interpret the \chandra\ results, since the most recent
\chandra\ calibration shows negligible cross-calibration flux
difference for \rosat\ PSPC data \citep{vikh05}.  For the hard
band, no single instrument has measured the range of fluxes necessary
for this calculation.  We instead use the \citetalias{more03} composite \logn\ curve for 2--10 keV, which uses bright-source data from
\xmmnewton\
and \asca\ with a total survey area of 71 deg$^{-2}$.

Because these \logn\ curves are for 0.5--2 keV and 2--10 keV, but we
wish to calculate intensities for 1--2 keV and 2--8 keV, we need to
determine how the source spectra vary with flux.  Following the
approach of \citetalias{more03}, we use a smooth function to
approximate the dependence of $\Gamma$ with flux. We use data obtained
with \rosat\ \citep{vikh95b} for the soft band (excluding stars which
give a negligible contribution), and \asca\ \citep{dell99} and
\chandra\ \citep{rosa02} for the hard band (Fig.\ \ref{figgam}), and
fit them with a scaled error function,
\begin{equation}
\Gamma(s)=a_0+a_1{\rm erf}\left(\log{\frac{S-a_2}{a_3}}\right)
\end{equation}
where $S$ is in units of $10^{-15}$ \flux.  For the fits we disregard
the different error bars on each data point to avoid biases. The best-fit parameter values are
$(a_0,a_1,a_2,a_3)=(0.59,1.7,-4.6,2.0)$ for 0.5--2 keV and
$(-16,18,-12,0.41)$ for 2--10 keV. Fig.\ \ref{figgam} shows these
fits; the results will not depend significantly on the details of
this approximation.

As mentioned above, we now integrate the sources detected in CDF
up to a certain cutoff flux, after which we integrate the \logn\ from
wider-area surveys.  For our purposes, the optimum upper flux cutoff
should be high enough so that we measure as much of the CXB signal as
possible using the CDF data itself (minimizing any cross-calibration
error), but low enough to limit the shot noise from our bright
sources.  Using the total sky coverage of each field (\S\
\ref{sourspec}) and using the above \logn\ curves, we find optimum
flux cutoffs of $5\times10^{-15}$ \flux\ for 0.5--2 keV sources and
$1.4\times10^{-14}$ \flux\ for 2--8 keV sources.  These give Poisson
errors corresponding to $\sim$5\% of the total CXB for the 1--2 and
2--8 keV bands, and still allow us to measure $\sim$50--60\% of the
total CXB using CDF data.

These cutoffs exclude several bright sources in each field.  We
exclude 5 sources in CDF-N and 4 in CDF-S for 0.5--2 keV, and 3 in
CDF-N and 2 in CDF-S for 2--8 keV.  The total aperture-corrected
fluxes of the sources fainter than these cutoffs, divided by the solid
angle of our $r=5$\arcmin\ regions (see \S\ \ref{calcdiff}) are given
in Table \ref{tblsrc}.  The CDF-S source intensity is $\sim$30\% lower
than that for CDF-N, consistent with the observed differences in the
\logn\ curves for those fields \citep[e.g.,][]{bran01b, rosa02}.

The next step is to obtain the background intensity for sources
brighter than our flux cutoffs, using the \logn\ distributions of
\citet{vikh95a} and \citetalias{more03}, and the $\Gamma$ as a
function of flux shown in Fig.\ \ref{figgam}.  Table \ref{tblsrc}
gives these total intensities, integrated to a maximum flux of
$10^{-11}$ \flux\ in the 0.5--2 keV and 2--10 keV bands.  There is
very little contribution ($<0.3$\%) from still brighter sources.  For
the purposes of obtaining a total CXB value, we average the resolved
intensities from these two fields.

To find the statistical errors in the bright source corrections, for
each energy band we calculate the total number of sources in bins of
flux, given the \logn\ curve and survey area.  The statistical errors
quoted in Table \ref{tblsrc} are simply the total Poisson
uncertainties in these source counts.  For bright source corrections
we also include include an additional error of $\pm$5\% to account for
any cross-calibration uncertainties, as in \citetalias{delu04}.

To verify that the total CXB value does not depend significantly on
the value of the flux cutoff, we measure the aperture-corrected
intensity of the combined source spectrum, plus the above
bright-source correction, for several flux cutoffs above our optiumum
values.  These intensities are shown in Fig.\ \ref{figcutoff}.  The
errors shown include the measurement error on the fluxes, as well as
the Poisson error in the measured fluxes and the bright source
correction.  Fig.\ \ref{figcutoff} shows that the total flux varies as
bright sources are included in the CDF spectrum, however these
differences are always within the errors.  Fig.\ \ref{figcutoff} also
shows that the errors on the total CXB with the increasing cutoff.

\begin{deluxetable*}{lccccccccc}
\tablecolumns{10}
\tabletypesize{\footnotesize}
\tablecaption{Summary of errors in unresolved intensity\label{tblerr}}
\tablewidth{6.5in}
\tablehead{\colhead{} & \colhead{} & 
\multicolumn{2}{c}{Count rate error} & 
\multicolumn{3}{c}{BG normalization error} &
 \colhead{PL $\Gamma$} & 
 \colhead{Calibration} &
 \colhead{Total} \\
\colhead{Subset} &
\colhead{Intensity} &
\colhead{Sky} &
\colhead{BG} &
\colhead{Systematic} &
\colhead{Sky} &
\colhead{BG} &
\colhead{error} &
\colhead{error} &
\colhead{error}\\
\colhead{(1)} &
\colhead{(2)}&
\colhead{(3)}&
\colhead{(4)}&
\colhead{(5)} &
\colhead{(6)}&
\colhead{(7)}&
\colhead{(8)}&
\colhead{(9)}&
\colhead{(10)}}
\startdata
\cutinhead{1--2 keV}
CDF-N VF &   0.93 &   0.063 &   0.09 &   0.09 &   0.028 &   0.041 &   0.018 &   0.028 &   0.17 \\
CDF-N F &   0.99 &   0.074 &   0.10 &   0.12 &   0.029 &   0.040 &   0.020 &   0.030 &   0.20 \\
CDF-N S &   1.20 &   0.062 &   0.10 &   0.11 &   0.027 &   0.038 &   0.014 &   0.036 &   0.19 \\
Average &   1.04 &   0.039 &   0.10 &   0.06 &   0.016 &   0.040 &   0.017 &   0.031 &   0.14 \\
\cutinhead{2--8 keV}
CDF-N VF &   3.52 &   0.61 &   0.95 &   1.72 &   0.52 &   0.78 &   0.49 &   0.11 &   2.43 \\
CDF-N F &   3.05 &   0.61 &   0.92 &   1.73 &   0.52 &   0.78 &   0.50 &   0.09 &   2.37 \\
CDF-N S &   3.58 &   0.50 &   0.87 &   1.63 &   0.49 &   0.73 &   0.58 &   0.11 &   2.24 \\
Average &   3.39 &   0.33 &   0.91 &   0.98 &   0.29 &   0.76 &   0.53 &   0.10 &   1.69 \\
\enddata
\tablecomments{Table columns are described in detail in \S\ \ref{sumerr}.}
\end{deluxetable*}

Together with our unresolved and resolved CDF fluxes, we use these
bright source corrections to calculate total CXB intensities averaged
across the fields, which are given in Table \ref{tblsrc}.  These
totals correspond to a CXB power law normalization (for $\Gamma=1.4$)
of 10.9 photons cm$^{-2}$ s$^{-1}$ keV$^{-1}$ sr$^{-1}$ at 1 keV, and resolved
fractions of $77\pm3$\% for 1--2 keV and $80\pm8$\% for 2--8 keV.  We
note that our resolved fractions are somewhat lower than those
determined by \citetalias{more03}, who used similar values for the
total CXB from the literature but found higher resolved fractions.
This may be due to the different \chandra\ calibration that they used
(\S\ \ref{aciscal}).  We stress that we have measured the unresolved
and a significant fraction of the resolved CXB intensities using the
same instrument, so there is little cross-calibration uncertainty
involved.

\begin{deluxetable}{lrr}
\tabletypesize{\footnotesize}
\tablecaption{Components of total CXB intensity\label{tblsrc}}
\tablewidth{3.4in}
\tablehead{
\colhead{} &
\colhead{1--2 keV} &
\colhead{2--8 keV}}
\startdata
Bright sources\tnm{a} & $2.09\pm0.10\pm0.10$ & $6.0\pm0.3\pm0.2$ \\
\cutinhead{CDF-N VF + F} 
All sources\tnm{b} & $3.05\pm0.11$ & $12.4\pm0.4$ \\
Sources below cutoff\tnm{c} & $1.81\pm 0.07\pm0.24$ & $8.0\pm 0.5\pm1.1$ \\
Resolved\tnm{d} & $ 3.89\pm0.30$ & $14.0\pm1.3$ \\
\cutinhead{CDF-N VF}
Unresolved & $0.93\pm0.17$  & $3.5 \pm 2.4$ \\
Total\tnm{e} & $4.83\pm0.36$ & $17.5\pm2.8$ \\
\cutinhead{CDF-N F}
Unresolved & $0.99 \pm 0.20$ & $3.1\pm2.4$ \\
Total\tnm{e} & $4.89\pm0.38$ & $17.0\pm2.7$ \\
\cutinhead{CDF-S}
All sources\tnm{b} & $2.88\pm0.11$ & $10.5\pm0.4$ \\
Sources below cutoff\tnm{c} & $1.12 \pm 0.05\pm0.28$ & $7.2\pm0.5\pm1.3$ \\
Resolved\tnm{d} & $ 3.21\pm0.33$ & $13.2\pm1.5$ \\
Unresolved & $1.20 \pm 0.19$ & $3.6\pm2.2$  \\
Total\tnm{e} & $4.41\pm0.38$ & $16.8\pm2.6$  \\
\cutinhead{Average}
Unresolved & $1.04\pm0.14$ & $3.4\pm1.7$ \\
Resolved\tnm{f} & $3.55\pm0.23$ & $13.6\pm1.0$ \\
Total\tnm{g} & $4.59\pm0.29$ & $17.0\pm2.0$ \\
\enddata
\tablecomments{Intensites are in units of $10^{-12}$ \intens.  All
  flux measurement errors include a 3\% ACIS calibration uncertainty.}
\tnt{a}{ Integrated contribution from \logn\ for
  sources brighter than $5\times10^{-15}$ \flux\ for 1--2 keV
  \citep{vikh95a}, and $1.4\times10^{-14}$ \flux\ for 2--8 keV
  \citep{more03}.  The first error is a 5\% cross-calibration
  uncertainty, the second is Poisson uncertainty in the average
  intensity due to the limited number of sources in the surveys.}
\tnt{b}{ Total intensity of all detected CDF sources within 5\arcmin\ of the
  aimpoint.}
\tnt{c}{ Total intensity of detected CDF sources within 5\arcmin\ of the
  aimpoint, fainter than
   $5\times10^{-15}$ \flux\ (1--2 keV) and $1.4\times10^{-14}$
  \flux\ (2--8 keV).  The first error corresponds to measurement error, and the second is Poisson uncertainty in the average intensity due to the
  limited number of source sampled.}
\tnt{d}{ Intensity of sources below the cutoff flux plus the integral of
  \logn\ above the cutoff.}
\tnt{e}{ Sum of the intensities of the measured unresolved
  component, CDF sources below the cutoff, and bright sources.}
\tnt{f}{ The average resolved intensity between the CDF-N and CDF-S
  fields.}
\tnt{g}{ The sum of the average resolved intensity and the average
  unresolved intensity.}
\end{deluxetable}

\begin{deluxetable*}{lrrr}
\tablecaption{Bright source corrections to total CXB\label{tblcorr}}
\tablewidth{14cm}
\tablehead{\colhead{} &
\colhead{Area (deg$^2$)} &
\colhead{Uncorrected} &
\colhead{Corrected}}
\startdata
\cutinhead{1--2 keV}
\citet{gend95}\tnm{a} & $0.67$ & $ 3.7\pm 0.2$ & $ 4.0\pm 0.3$ \\
\citet{geor96} & $1.40$ & $ 4.8\pm 0.2$ & $ 5.0\pm 0.3$ \\
\citet{chen97}\tnm{b} & $0.22$ & $ 4.3\pm 0.2$ & $ 4.8\pm 0.4$ \\
\citet{miya98} & $0.30$ & $ 4.0\pm 0.3$ & $ 4.4\pm 0.4$ \\
\citet{parm99} & $0.50$ & $ 4.3\pm 0.2$ & $ 4.7\pm 0.3$ \\
\citet{vecc99} & $0.67$ & $ 4.6\pm 0.2$ & $ 4.9\pm 0.3$ \\
\citet{lumb02}\tnm{c} & $1.57$ & $ 3.5\pm 0.1$ & $ 4.5\pm 0.2$ \\
\citet{mark03}\tnm{d} & $0.07$ & $ 4.2\pm 0.4$ & $ 4.9\pm 0.6$ \\
This work\tnm{e} & $0.06$ & $2.5\pm0.1$ & $4.6\pm0.2$ \\

\cutinhead{2--8 keV}
\citet{gend95}\tnm{a} & $0.67$ & $14.1\pm 0.6$ & $15.1\pm 0.9$ \\
\citet{chen97}\tnm{b} & $0.22$ & $14.9\pm 1.2$ & $15.4\pm 1.2$ \\
\citet{miya98} & $0.30$ & $15.1\pm 1.6$ & $16.6\pm 1.9$ \\
\citet{vecc99} & $0.67$ & $17.5\pm 0.8$ & $18.6\pm 1.1$ \\
\citet{delu04} & $5.50$ & $17.9\pm 0.8$ & $18.3\pm 0.9$ \\
This work\tnm{e} & $0.06$ & $10.6\pm1.7$ & $16.6\pm1.9$ \\
\enddata \tablecomments{Intensities are in units of $10^{-12}$
\intens.}  
\tnt{a}{ Using their CXB normalization for the 1--7 keV power law fit.}  
\tnt{b}{ Using their CXB normalization for the joint \rosat/\asca\ fit.}  
\tnt{c}{ In the 1--2 keV band, we correct for bright sources
with $S>2\times10^{-14}$ \flux, which were excluded from their spectral
analysis.}
\tnt{d}{ Using their normalization for the 1--7 keV power law fit for
  4 fields.  The 1--2 keV flux is corrected downward by 7\% to account
for recent ACIS calibration (\S\ \ref{aciscal}). }
\tnt{e}{ We correct for bright sources with
  $S>5\times10^{-15}$ \flux\ (0.5--2 keV) and $S>1.4\times10^{-14}$
  \flux\ (2-8 keV), see text. The area here is the full solid angle
  covered by $5'$ circles in all pointings.}
\end{deluxetable*}

\subsubsection{Summary of errors}\label{sumerr}
Our calculation of the CXB intensity involves a number of different
sources of uncertainty.  For clarity, we give here a summary of these
errors and their propagation.  We begin with the errors in the
unresolved intensity, which are given in detail in Table
\ref{tblerr}, described below.

\begin{enumerate}
\im Columns (1) and (2) of Table \ref{tblerr} give the subset of the
  data and the corresponding unresolved intensity, corrected for
  scattered source flux (\S\ \ref{psfscatter}).  All units are
  $10^{-12}$ \intens.

\im Columns (3) and
  (4) give the uncertainty due to statistical errors in the sky and
  background count rates, respectively, for the 1--2 keV or 2--8 keV band.  

\im Columns (5)--(8) give errors due to uncertainty in the stowed
  background normalization.  Column (5) gives the error due to the 2\%
  systematic uncertainty in the stowed background spectral shape (\S\
  \ref{backnorm}, while columns (6) and (7) give errors due to
  statistical uncertainties in the 9--12 keV sky and background count
  rates, which are 0.7--0.9\% and 0.5--0.6\%, respectively.  The
  corresponding errors on the unresolved intensity shown here reflect the
  fact that the stowed background count rate is $\simeq$5 and 25 times larger
  than the unresolved signal in the 1--2 keV and 2--8 keV bands,
  respectively.

\im Column (8) gives the variation in intensity for the
  range of power law photon index $\Gamma=$1.1--2.0. 

\im Column (9) gives the 3\% error on the ACIS flux calibration. 

\im Column (10) gives the total error on the unresolved intensity,
  calculated assuming all the above errors are independent.

\end{enumerate}

Errors on the average unresolved CXB are calculated by propagation of
these errors.  We note, however, that the stowed background data are
essentially identical for each subset, the power law index is expected
not to vary between subsets, and the ACIS calibration error is not
statistical in nature.  Therefore the errors in columns (4), (7), (8),
and (9) are not independent between subsets, and so do not decrease
when averaging the subsets together.

For the total CXB intensity, our calculation has three components as
listed in Table \ref{tblsrc}:  (1) the unresolved CXB
described above, (2) resolved sources in the CDF below our
cutoff fluxes, and (3) the bright source correction.  The errors in
each component are described below.

\begin{enumerate}
\im Errors in the unresolved intensity are as above.
\im For resolved sources in the CDF, we include two errors in Table
\ref{tblsrc}.  The first gives the measurement error, including
statistical count rate errors and the 3\% ACIS flux calibration
uncertainty.  Errors in the stowed background, which are important for
the unresolved intensity, are negligible here because the large ratio
of source to background counts.  The second, and larger, error in the
resolved source intensity is the Poisson uncertainty in estimating the
average CXB intensity across the sky, as described in \S\ \ref{sourcedist}.  In
calculating the average of these values between fields, we propagate
all these errors except the 3\% calibration error.
\im For the bright source correction, we also give two sources of
error in Table \ref{tblsrc}.  The first is the 5\% cross-calibration
uncertainty.  The second is the Poisson error in the average
intensity, as above.
\end{enumerate}

For the total CXB intensity, we treat as independent the above errors
except the 3\% ACIS flux calibration uncertainty.

\subsubsection{Cosmic variance}

We note that the resolved and total CXB intensities are somewhat
higher for CDF-N than for CDF-S, which may be evidence for significant
large-scale structure variations between fields
\citep[e.g.,][]{gill03, gill05,
barg03, yang03}.  Here we estimate the uncertainty in our average
total CXB due to cosmic variance.  For a population of sources with a
two-point angular correlation function $w(\theta)$, the variance in the
number counts in a field of solid angle $\Omega$ can be estimated by
 \citep[see Eqn. (45.6) of][]{peeb80}:
\begin{equation}
\sigma_{\Omega}^2=\iint{w({\bf \theta_1}-{\bf \theta_2}){\rm
    d}\Omega_1{\rm d}\Omega_2} \label{eqncor}.
\end{equation}
Clustering of extragalactic X-ray sources can be described by
$w(\theta)=(\theta/\theta_0)^{1-\gamma}$, with $\gamma\simeq1.8$ and
$\theta_0=$4\arcsec--10\arcsec \citep[e.g.,][]{vikh95c,basi05}.
Evaluating Eqn. (\ref{eqncor}), this gives expected cosmic variance
for each of our two 5\arcmin\ radius CDF regions of $\sim$20--30\%.
We use CDF data for 50--60\% of our total CXB intensity
estimate while the rest comes from wide-field surveys, so the above
cosmic variance corresponds to $\sim$10--20\% error in our total CXB values.

For wider fields, \citet{kush02} found that in the 2--10 keV band, the
variation in CXB intensity between \asca\ GIS fields of radius
20\arcmin\ was $\simeq$6\%.  This is consistent with only Poisson
fluctuations but may be difficult to distinguish from cosmic
variance, which given the above models above can be as low as $\sim$10\%
for a field of 20\arcmin\ radius.  We conclude that the possible error in our
total CXB values due to large-scale structure is difficult
to predict, but is likely in the range $\sim$10--20\%.  Because of the
uncertainty in this error, we do not include it in our final results.

\subsubsection{Bright source correction for other works}
We next compare our total CXB intensities to the earlier works listed
in \S\ \ref{intro}.  Many studies of the total CXB give only a power
law slope and normalization for the CXB spectrum; for these
measurements, we include only the error on the normalization for
simplicity.  For uniformity, do not include cross-calibration
uncertainties.  To compare directly with our results, for surveys with
limited sky coverage we have applied a bright source correction
exactly as in \S\ \ref{sourcedist}.  The upper ``cutoff'' flux from which to
integrate the \logn\ is calculated from the area of each survey, as
the flux at which the survey should have on average one source.

These corrections, and the uncertainties calculated including the
shot noise are given in Table \ref{tblcorr}, and the resulting total
CXB intensities are shown in Fig.\ \ref{figtot} (note that the values
and uncertainties differ somewhat from those in Fig.\ 3 of
\citetalias{more03}.  We have reduced the 1--2 keV value of
\citet{mark03} by 7\% to account for changes in the ACIS calibration
(\S\ \ref{aciscal}).  Fig.\ \ref{figtot} shows that our total CXB values are in
good agreement with previous measurements.

\subsubsection{Error bar comparison to other works}                                               
Our uncertainty for the 2--8 keV total CXB flux is considerably higher
than the 4\% error (68\%) quoted by \citetalias{delu04} for their
\xmmnewton\ measurement (Fig.\ \ref{figtot}).  While the \xmmnewton\
EPIC MOS has several times higher ratio of the sky signal to the
detector background in this band than \chandra\ \mbox{ACIS-I}, its background
is much more variable, so one would expect the resulting CXB
accuracies to be comparable.  However, our and \citetalias{delu04} error bars cannot
be directly compared, because the \xmmnewton\ error does not include
an analog of our systematic uncertainty of the detector background
modeling, which dominates our 2--8 keV error bar. While DM04 made a
considerable effort to quantify their background uncertainties and
concluded that they are negligible, they may in fact be significant.
\citetalias{delu04} did not present the scatter of the source-free 2--8 keV
background rates in their individual pointings that remained after
their flare filtering, which might give a direct measure of this
uncertainty.  However, \citet{neva05}, using a more aggressive flare
filtering (which on average discarded 35\% of the \xmmnewton\ MOS
data), still observed a 6\%  rms scatter of the 2--4 keV count
rate in their blank-sky fields. This would already correspond to a
$\sim 20$\% scatter in the measured CXB flux. When such systematic
uncertainties are accounted for, the \chandra\ and \xmmnewton\ error
bars should become comparable.  

The error of the \mbox{ACIS-S} result of \citet{mark03} is
significantly larger than ours; it was dominated by the statstical error
of the short (11 ks) dark Moon observation used as a background in
that work.

\subsection{Nature of the unresolved CXB}

The unresolved X-ray background intensity represents the
integrated flux of all types of X-ray sources below the CDF flux
limits, plus any truly diffuse component.  Here we examine
if it can be accounted for by any known source populations.

\defcitealias{baue04}{B04}
\defcitealias{rana03}{R03}

\subsubsection{Extrapolation of \logn\ to lower fluxes}
Here we test if the observed \logn\ for the CDF, extrapolated to
lower fluxes, can account for the unresolved CXB intensity.  
 \citet[][hereafter   B04]{baue04} presents a single power law fit to the 
\logn\ for all CDF sources at $S <10^{-15}$ \flux\ (the thick black line in Fig.\
\ref{figextr}).  We extrapolate this curve from the CDF
flux limits down to $10^{-18}$ \flux\ (0.5--2 keV) and $10^{-17}$ 
\flux\ (2--8 keV), and convert fluxes from 0.5--2 to 1--2 keV using
our unresolved best-fit $\Gamma=1.5$. 
The integrated intensities are given in Table \ref{tblextr}.  These values fall
well below our observed unresolved intensities (at the level of
6$\sigma$ for 1--2 keV and 1.5$\sigma$ for 2--8 keV).  If the CXB is
completely due to point sources, this is direct
evidence for a steepening of the $\log{N}/\log{S}$ curve for low flux.

\subsubsection{A new population of point sources?}
We next examine how contributions from separate populations of point
sources could give the unresolved CXB intensity.  There is
considerable evidence that within a factor of 10 below the CDF flux
limits, the number density of sources should become dominated by
starburst and normal galaxies, rather than AGNs.  \citet{miya02}
performed a fluctuation analysis on unresolved parts of the 1 Ms CDF
images, and found a possible upturn in the distribution just below the
limiting CDF fluxes.  \citet[][hereafter R03]{rana03} examined the
relationship between radio and X-ray emission from star-forming
galaxies, and predicted that such galaxies would dominate the CXB
sources at fluxes $\lesssim$$10^{-17}$ \flux.  Most directly,
\citetalias{baue04} produced separate \logn\ curves for AGNs and
galaxies, selected by their optical and X-ray properties, and showed
that the galaxy numbers rise steeply and may begin to dominate at low flux.

\begin{figure}[t]
\epsscale{1.1}
\plotone{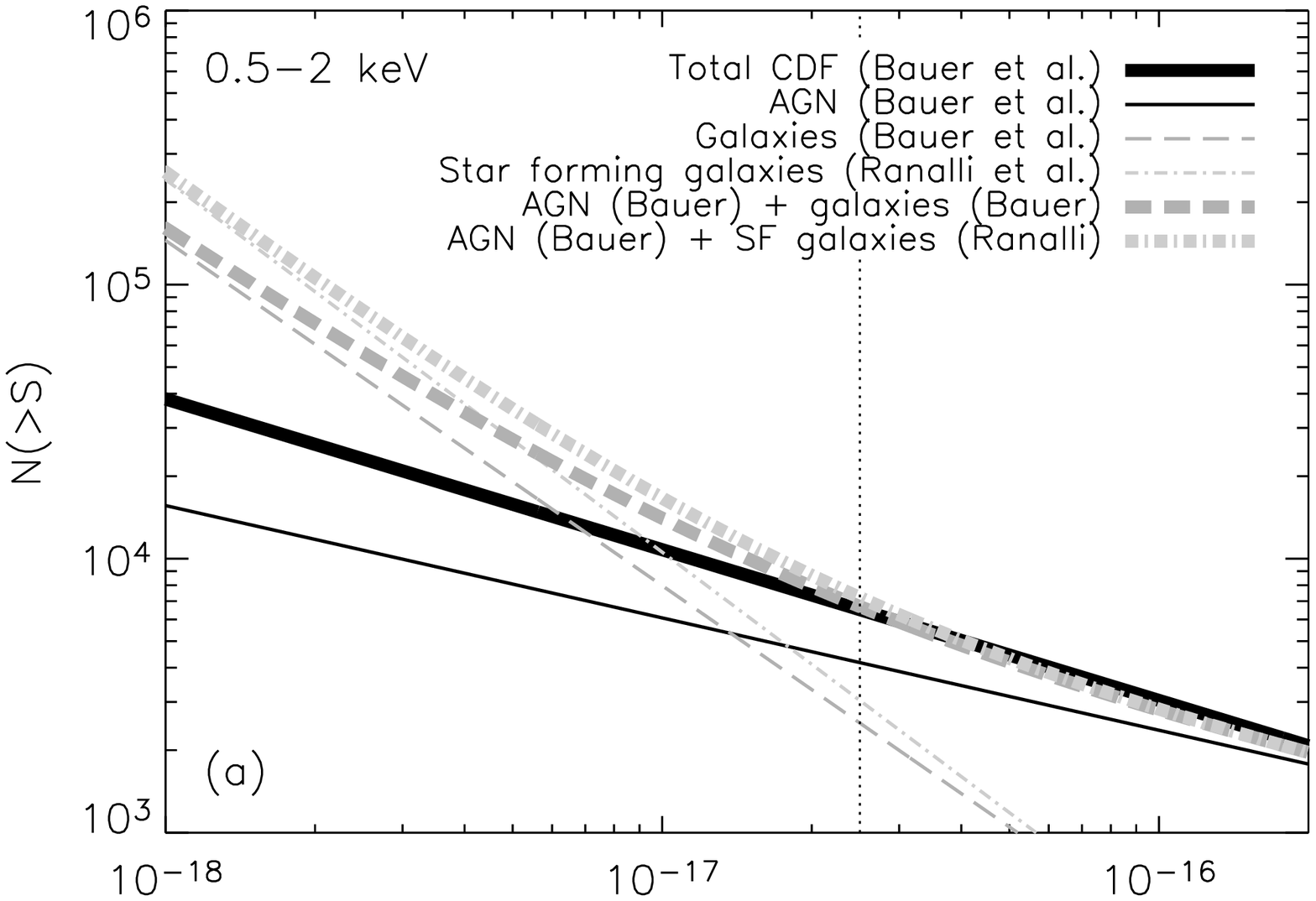}
\vskip-0.25in
\plotone{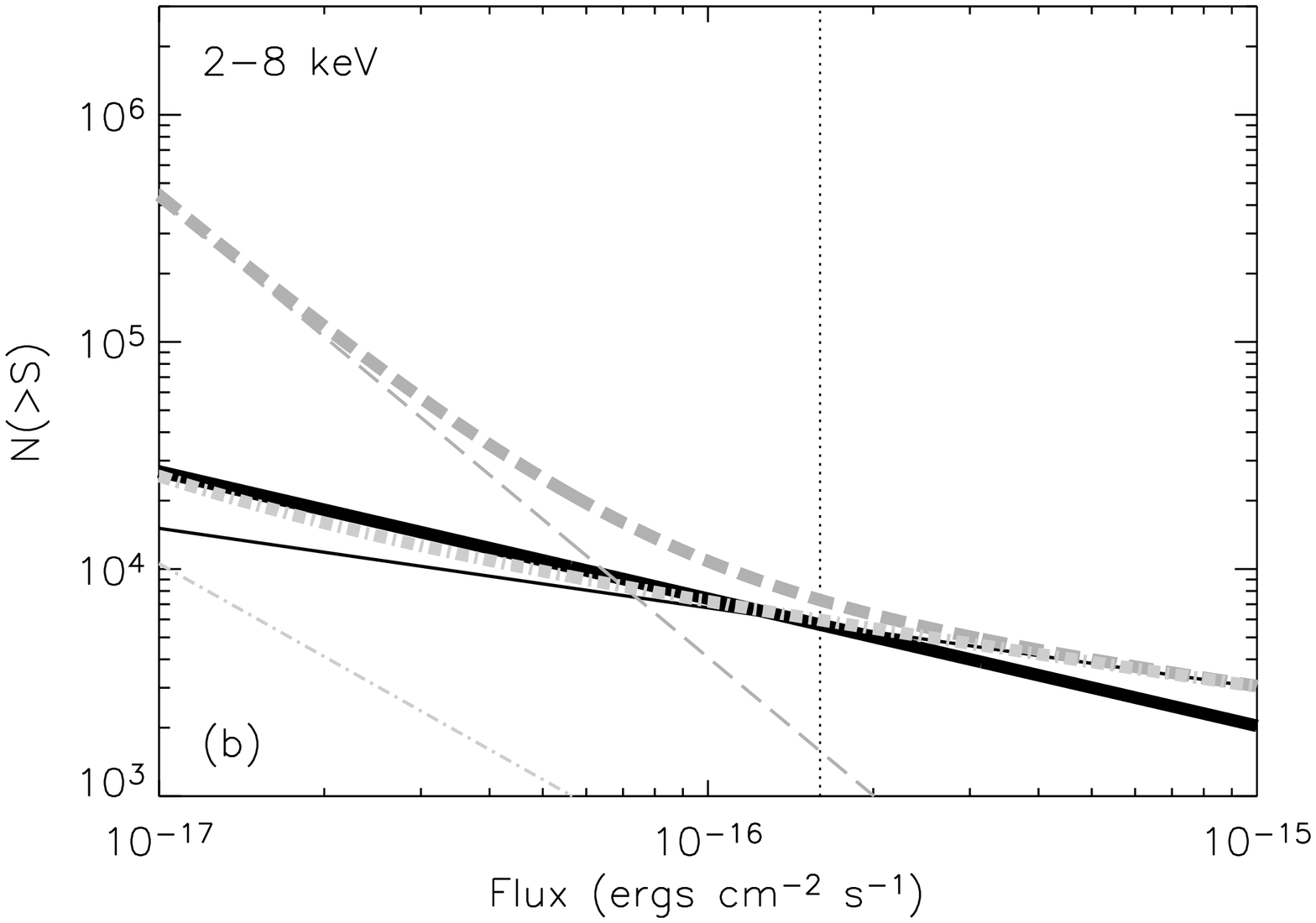}
\caption{Best-fit \logn\ distributions for faint sources, for 0.5--2
   (top) and 2--8 keV (bottom).  The distribution for all CDF detected
   sources, as well as AGNs and galaxies individually, are taken from
   \citetalias{baue04}.  An estimate for star-forming galaxies is from
   \citetalias{rana03}.  The sums of the \citetalias{baue04} AGNs and the
   separate galaxy distributions are also shown.  Vertical dotted
   lines show the \citetalias{alex03} CDF-N flux limits. \label{figextr}}
\end{figure}

Here we use the \citetalias{baue04} best-fit separate \logn\ curves
for AGNs and galaxies (shown in Fig.\ \ref{figextr}) and again
extrapolate them from the CDF flux limits down to $10^{-18}$ \flux\
(0.5--2 keV) and $10^{-17}$ \flux\ (2--8 keV).  We use their
classification that is ``optimistic'' for galaxies (and
correspondingly ``pessimistic'' for AGNs), which gives the maximum
number of objects at low flux.  To convert 0.5--2 keV to 1--2 keV
fluxes, we again use $\Gamma=1.5$ for simplicity.  The extrapolated
intensities are given in Table \ref{tblextr}.

For 1--2 keV, extrapolating to $S=10^{-18}$ \flux, we find a total
integrated intensity $(4.0^{+0.7}_{-0.5})\times10^{-13}$ \intens,
still short of the observed signal.  The predicted flux distribution
of star-forming galaxies from \citet{rana03} would give a slightly
larger signal (see Table \ref{tblextr}), but still requires a
significant population of sources with $S<10^{-18}$ \flux.  Thus, the
\logn\ may continue to rise steeply down to fluxes more than an order
of magnitude fainter than the CDF flux limits.

For the 2--8 keV band, extrapolating down to $S=10^{-17}$ \flux, the
contribution from \citetalias{baue04} AGNs is well below the observed
value, but the contribution from galaxies is much larger.  However,
the large errors in the galaxy distribution, due to a small number of
sources at low fluxes, make it difficult to place meaningful
constraints on the contribution from galaxies.  If the best-fit
\citetalias{baue04} galaxy $\log{N}/\log{S}$ curve is formally
extrapolated to $10^{-17}$ \flux, it can account for the full observed
unresolved intensity.  We note however that the distribution for
star-forming galaxies predicted by \citetalias{rana03} is
significantly flatter, and would produce an intensity lower than that
observed.  Thus, the nature of the unresolved source population for
$E>2$ keV remains unclear.

Recently, \citet{wors06} performed X-ray stacking of optical
sources in the CDF fields that have no detected X-ray
counterparts.  They found that these sources can contribute
$\sim$15-20\% of the total CXB in the 1--4 keV band, providing
evidence that X-ray point sources produce the majority
of the unresolved 1--4 keV CXB.

\subsubsection{WHIM?}
The putative ``warm-hot'' diffuse intergalactic medium with
$T=10^5-10^7$~K \citep[WHIM,][]{cen99, dave01}
should contribute to our unresolved flux at $E<2$ keV.  In the 1--2
keV band, the expected WHIM signal is very low, 30--50 times below our
unresolved brightness \citep{phil01}. However, the WHIM
becomes much brighter below 1 keV. A future paper will attempt to
compare the WHIM predictions with our unresolved CXB spectrum at low
energies.

\begin{deluxetable*}{lrr}
\tablecaption{Extrapolated intensities from $\log{N}/\log{S}$
  distributions \label{tblextr}}
\tablehead{
\colhead{} &
\colhead{1--2 keV} &
\colhead{2--8 keV}}
\startdata
Observed (CDF-N VF) & $10.5\pm1.4$ & $35 \pm 17$ \\
Total CDF \citep{baue04} & $1.5\pm0.1$ & $8.2^{+3.1}_{-2.6}$ \\
AGN \citep{baue04} & $0.60^{+0.06}_{-0.05}$ & $4.3^{+2.5}_{-2.2}$ \\
Galaxies \citep{baue04} & $3.9^{+0.6}_{-0.5}$ & $92^{+60}_{-92}$ \\
Star-forming galaxies \citep{rana03} & 4.7 & 2.5 \\ 
AGN + galaxies \citep{baue04} & $4.5^{+0.08}_{-0.07}$ &
  $96^{+60}_{-92}$ \\
AGN \citep{baue04} + SF galaxies \citep{rana03} & 5.3 & 6.8 \\
\enddata
\tablecomments{Intensities in $10^{-13}$ \intens.  The \logn\ curves
  are extrapolated from the CDF-N flux limits (\S\ \ref{sourexcl})
  down to $10^{-18}$ and $10^{-17}$ \flux\ for
  the 0.5--2 and 2--8 keV bands, respectively.}
\end{deluxetable*}

\subsubsection{Galaxy groups?}
Finally, we compare our 1--2 keV unresolved signal with surface
brightness of galaxy groups, which, according to some estimates
\citep[see e.g.,][for a high-end prediction]{brya01}, may together
contribute at this level to the CXB. For a typical group with average
$kT\simeq 2$ keV at $z\simeq 0.1$, the 1--2 keV surface brightness at
$r_{200}$, the radius of overdensity 200 times the critical density
(i.e., far in its outskirts) would be of order $3\times 10^{-13}$
\intens \citep[e.g.,][]{vikh05}. This is 3--4 times below our observed
flux.  The surface brightness in the core of a group is 2.5--3 orders
of magnitude higher. The radius $r_{200}=7-10'$ at that redshift;
groups closer than that would be big and bright enough to be easily
seen in \rosat\ surveys, and none is seen near either of the two CDF
fields in \rosat\ pointed or all-sky data (nor in the \chandra\
pointings immediately around the CDF-S field). Thus, the diffuse flux
is not because either field is in the outskirts of a nearby group. At
the same time, more distant galaxy groups at $z\simeq 0.1-0.7$ would
be easily detected in these deep \chandra\ images and excluded by
us. Indeed, our excluded extended sources in Fig.\ 1 are such objects
\citep{giac02}. The surface brightness of still more distant groups
and clusters quickly decreases with redshift because of cosmological
dimming.  For a quantitative estimate of the cumulative brightness of
undetected distant groups and clusters One might use the cluster
\logn\ \citep[e.g.,][]{rosa02a} for a quantitative estimate of the
cumulative brightness of undetected distant groups and clusters, but
it is already obvious that groups cannot contribute significantly to
our unresolved 1--2 keV flux.

\section{Summary}
We measure the absolute intensity of the unresolved cosmic X-ray
background, after the exclusion of the sources detected in the deepest
\chandra\ observations, the \chandra\ Deep Fields North and
South.   We find  significant residual CXB intensity in the
0.5--1 and 1--2 keV bands, and a marginal (2$\sigma$) signal in
2--8 keV band.  There is unlikely to be any significant contamination of the
signal from ACIS instrumental background.

We find unresolved intensities of $(1.04\pm0.14)\times10^{-12}$ \intens\
and $(3.4\pm1.7)\times10^{-12}$ \intens\ in the 1--2 and 2--8 keV
bands.  These values are much higher than the extrapolation of the
observed \logn\ curve down to lower fluxes, suggesting an upturn of
the distribution below the CDF flux limits, or a truly diffuse
component.  Combining these results with the total contribution from
point sources from CDF and wider-field CXB surveys, we obtain
total CXB intensities of $(4.6 \pm 0.3) \times10^{-12}$ and $(1.7 \pm
0.2) \times10^{-11}$ \intens\ for the 1--2 and 2--8 keV bands,
respectively.  This implies resolved fractions of $77\pm3$\% in 1--2 keV and
$80\pm8$\% in 2--8 keV, lower than some previous claims.

\nocite{mars80,mcca83,ueda99,gore69,geor96,chen97,parm99,kunt01,revn03,revn04}



\begin{acknowledgements}
We thank A. Viklhinin, C. Jones, W.R. Forman, S.S. Murray, R. Narayan,
 S. Virani, J.P. Ostriker, and A. Finoguenov for useful discussions,
 and the referee, F. Bauer, for helpful comments.  RCH was supported
 by a NASA GSRP Fellowship NNG05GO20H, NSF grant AST-0307433, and
 \chandra\ grant G03-4176A, and MM by NASA contract NAS8-39073 and
 \chandra\ grant G04--5152X.
\end{acknowledgements}


\clearpage

\newpage

\clearpage

\clearpage

\end{document}